\documentclass[submission, Phys]{SciPost}
\usepackage{amsmath,amssymb,amsfonts,mathtools}
\usepackage{amsthm}
\usepackage{nicefrac}
\usepackage{graphicx}
\usepackage{hhline}
\usepackage[utf8]{inputenc}
\usepackage{xcolor}

\usepackage{hyperref}
\hypersetup{
	hypertexnames=false,  
    colorlinks=true,       
    linkcolor=blue,          
    citecolor=blue,       
    filecolor=magenta,      
    urlcolor=blue          
}

\usepackage{soul} 

\newcommand{\beq}{\begin{equation}}
\newcommand{\eeq}{\end{equation}}
\newcommand{\nn}{\nonumber\\}
\newcommand{\eq}[1]{Eq.~(\ref{eq:#1})}
\newcommand{\Eq}[1]{Equation~(\ref{eq:#1})}
\newcommand{\eqs}[2]{Eqs.~(\ref{eq:#1}) and (\ref{eq:#2})}

\newcommand{\fref}[1]{Fig.~\ref{fig:#1}}

\newcommand{\sref}[1]{Sec.~\ref{sec:#1}}
\newcommand{\srefs}[2]{Secs.~\ref{sec:#1} and~\ref{sec:#2}}
\newcommand{\Sref}[1]{Section~\ref{sec:#1}}

\newcommand{\eql}[1]{\label{eq:#1}}
\newcommand{\figl}[1]{\label{fig:#1}}
\newcommand{\secl}[1]{\label{sec:#1}}

\newcommand{\ket}[1]{\vert#1\rangle}
\newcommand{\bra}[1]{\langle#1\vert}
\newcommand{\ip}[2]{\langle#1\vert#2\rangle}
\newcommand{\me}[3]{\langle#1\vert#2\vert#3\rangle}

\newcommand{\Real}{{\rm Re}}
\newcommand{\Imag}{{\rm Im}}

\def\q{{\bf q}}
\def\R{{\bf R}}
\def\bpartial{\boldsymbol{\partial}}
\def\rr{{\bf r}}  
\def\kk{{\bf k}}
\def\v{{\bf v}}
\def\m{{\bf m}}
\def\d{{\bf d}}

\def\A{{\bf A}}
\def\AA{\boldsymbol{\mathcal{A}}}

\def\D{{\bf D}}

\def\M{{\bf M}}
\def\P{{\bf P}}
\def\S{{\bf S}}
\def\0{{\mathbf{0}}}

\def\Emat{\mathcal{E}}
\def\Mmat{\mathcal{M}}

\def\bnk{_{n\kk}}

\def\blk{_{l\kk}}

\def\en{\varepsilon_n}
\def\el{\varepsilon_l}

\def\m{{\bf m}}
\def\w{\omega}
\def\ww{\tilde\w}

\def\bnabla{{\boldsymbol\nabla}}
\def\doteq{\overset{\boldsymbol{.}}{=}}
\def\btau{{\boldsymbol\tau}}

\def\flipab{(a\leftrightarrow b)}

\def\bxi{\boldsymbol{\xi}}

\def\EDbar{\bar{\text{E}}1}
\def\MDbar{\bar{\text{M}}1}
\def\EQbar{\bar{\text{E}}2}

\def\T{\mathcal{T}}
\def\P{\cal{P}}

\def\talpha{\tilde\alpha}
\def\calpha{\check\alpha}

\begin{document}

\begin{center}{\Large \textbf{ Multipole theory of optical spatial
    dispersion in crystals }}\end{center}

\begin{center}
\'Oscar Pozo\textsuperscript{1},
Ivo Souza\textsuperscript{1,2}
\end{center}

\begin{center} {\bf 1} Centro de F{\'i}sica de Materiales, Universidad
del Pa{\'i}s Vasco, 20018 San Sebasti{\'a}n, Spain
\\
{\bf 2} Ikerbasque Foundation, 48013 Bilbao, Spain
\end{center}

\begin{center}
\today
\end{center}


\section*{Abstract}

{\bf Natural optical activity is the paradigmatic example of an effect
  originating in the weak spatial inhomogeneity of the electromagnetic
  field on the atomic scale. In molecules, such effects are well
  described by the multipole theory of electromagnetism, where the
  coupling to light is treated semiclassically beyond the
  electric-dipole approximation.  That theory has two shortcomings: it
  is limited to bounded systems, and its building blocks --~the
  multipole transition moments~-- are origin dependent.  In this work,
  we recast the multipole theory in a translationally-invariant form
  that remains valid for crystals. Working in the independent-particle
  approximation, we introduce ``intrinsic'' multipole transition
  moments that are origin independent and transform covariantly under
  gauge transformations of the Bloch eigenstates.  Electric-dipole
  transitions are given by the interband Berry connection, while
  magnetic-dipole and electric-quadrupole transitions are described by
  matrix generalizations of the intrinsic magnetic moment and quantum
  metric. In addition to multipole-like terms, the response of
  crystals at first order in the wavevector of light contains
  band-dispersion terms that have no counterpart in molecular
  theories. The full response is broken down into magnetoelectric and
  quadrupolar parts, which can be isolated in the static limit where
  electric and magnetic fields become decoupled. The rotatory-strength
  sum rule for crystals is found to be equivalent to the topological
  constraint for a vanishing chiral magnetic effect in equilibrium,
  and the formalism is validated by numerical tight-binding
  calculations.}

\vspace{10pt}
\noindent\rule{\textwidth}{1pt}
\tableofcontents\thispagestyle{fancy}
\noindent\rule{\textwidth}{1pt}
\vspace{10pt}

\section{Introduction}
\secl{intro}

As the wavelength of optical radiation is large compared to atomic
dimensions, the interaction of light with matter is generally well
described by taking the long-wavelength limit (electric-dipole
approximation).  In that approximation, the response of the medium to
an electromagnetic perturbation is treated as local in space.  When
nonlocality is taken into account, the response acquires a dependence
on the wavevector $\q$ of light, and this is known as spatial
dispersion~\cite{landau-book84,melrose-book91}.

Although the effects of spatial dispersion can often be treated as
small corrections, they are significant in that they lead to
qualitatively new phenomena. One example is natural optical
activity~\cite{landau-book84}, whose most familiar manifestation is
the rotation of the plane of polarization of linearly-polarized light
travelling through chiral molecules in solution.  Lesser-known
manifestations of spatial dispersion include gyrotropic birefringence
and nonreciprocal directional
dichroism~\cite{hornreich-pr68,graham-pmb92}; these are
magneto-optical effects that occur in acentric magnetic materials
without requiring a net magnetization.

Because of the fundamental and industrial importance of chiral
molecules, molecular quantum theories of natural optical activity
--~and, by extension, of other spatially-dispersive optical effects~--
have been developed over many decades, on the basis of the multipole
theory of
electromagnetism~\cite{condon-rmp37,barron-book04,raab-book05,norman-book18}.
This has led to the development, starting in the mid 1990s, of several
\textit{ab initio} methods for calculating optical rotatory dispersion
and natural circular dichroism spectra of molecules. Some of those
methods rely on sum-over-states formulas~\cite{pedersen-cpl95}; in
others, the explicit summation over a truncated set of excited states
is avoided using either static-limit~\cite{polavarapu-molphys97} or
finite-frequency~\cite{varsano-pccp09} linear-response schemes, or
real-time propagation approaches~\cite{yabana-pra99,varsano-pccp09}
(see
Refs.~\cite{polavarapu-chirality02,autschbach-review11,mattiat-hca21}
for reviews).

By comparison, there have been relatively few attempts to formulate
bulk theories of optical spatial
dispersion~\cite{moffitt-jcp56,kato-jpsj73,agranovich-book84},
particularly within the one-electron band
picture~\cite{natori-jpsj75,zhong-prb93,jonsson-prl96,malashevich-prb10,zhong-prl16,ma-prb15,gao-prl19,duff-prb22}.
As a result, only a small number of \textit{ab initio} calculations of
natural optical
activity~\cite{zhong-prb93,jonsson-prl96,hidalgo-prb09,tsirkin-prb18,rerat-jctc21}
and of nonreciprocal directional
dichroism~\cite{yokosuk-npjqm20,park-npjqm22} have been carried out
for crystals.  Such effects provide valuable information about broken
structural and magnetic symmetries, and their study in novel
bulk~\cite{arima-jpcm08,yokosuk-npjqm20,park-npjqm22} and
quasi-two-dimensional~\cite{kim-natnano16,suarez-2dmater17,stauber-prl18,chang-prb22,han-advmater23}
materials calls for improved theoretical descriptions.

In this work, we develop a microscopic theory of optical spatial
dispersion in crystals that is firmly rooted in the molecular
multipole theory.  We work in the independent-particle approximation
neglecting local-field effects~\cite{jonsson-prl96}, and focus on the
electronic response with frozen ions. We proceed by evaluating the
optical conductivity at first order in $\q$, including both interband
and intraband contributions, and arrive at a sum-over-states
expression in terms of well-defined multipole-like transition moments.

To set the stage, let us introduce the multipole transition moments as
defined in the standard molecular
theory~\cite{raab-book05,norman-book18}.  The electric dipole (E1)
appears at leading order in the multipole expansion, followed at the
next order by the magnetic dipole (M1) and electric quadrupole (E2).
These are the needed ingredients to describe natural optical activity,
gyrotropic birefringence, and nonreciprocal directional dichroism. In
the independent-particle approximation, they take the form
\begin{subequations}
\eql{E1-M1-E2}
\begin{align}
\d_{nl}&=-e\me{\phi_n}{\rr}{\phi_l}\,, \eql{E1}\\
\m_{nl}&=-\frac{e}{2}\me{\phi_n}{\rr\times\v}{\phi_l}\,,\eql{M1}\\
q^{ab}_{nl}&=-e\me{\phi_n}{r_ar_b}{\phi_l}\,,\eql{E2}
\end{align}
\end{subequations}
where $\phi_n(\rr)$ and $\phi_l(\rr)$ are occupied and empty energy
eigenstates of the molecule, respectively, and $-e$ is the electron
charge.  (The M1 transition moment also has a spin part; we omit it
for now, but it will be included later.)

In trying to extend the multipole theory to periodic crystals, one is
faced with the problem of how to define the transition moments when
the molecular orbitals $\phi_n(\rr)$ are replaced by Bloch eigenstates
$\psi\bnk(\rr)=e^{i\kk\cdot\rr}u\bnk(\rr)$, given that the matrix
elements in \eq{E1-M1-E2} involve the nonperiodic position operator
$\rr$.  For E1 transitions between nondegenerate bands $n$ and $l$,
there is a well-known prescription, namely
$\d_{nl}(\kk)=-e\ip{u\bnk}{i{\boldsymbol\partial}_\kk
  u\blk}$~\cite{blount-ssp62,vanderbilt-book18}.

The situation is less clear when it comes to defining M1 and E2
transition moments in the Bloch representation.  Already for
molecules, their definitions in \eq{E1-M1-E2} are somewhat
problematic, as they give values that change under a rigid shift of
the coordinate system. Molecular properties should be origin
independent, and for spatially-dispersive optical coefficients that is
generally ensured by a cancellation between the origin dependences of
different terms of the same order in the multipole
expansion~\cite{barron-book04,raab-book05,norman-book18,pedersen-cpl95}.
This is not entirely satisfactory from a formal standpoint, and
moreover it leads to slightly origin-dependent numerical results,
because the cancellation is not exact for incomplete basis
sets~\cite{pedersen-cpl95,norman-book18}.

In our independent-particle formulation, the optical conductivity at
first order in $\q$ is written in terms of ``intrinsic'' multipole
transition moments $\EDbar$, $\MDbar$, and $\EQbar$ that are origin
independent and well defined for both molecules and crystals.  For
molecules, these modified transition moments take the form
\begin{subequations}
\begin{align}
\bar\d_{nl}&=
-e\me{\phi_n}{\rr-\left(\bar\rr_n+\bar\rr_l\right)/2}{\phi_l}\,,
\eql{E1bar-mol}\\
\bar\m_{nl}&=
-\frac{e}{2}\me{\phi_n}
{\left[\rr-\left(\bar\rr_n+\bar\rr_l\right)/2\right]\times\v}{\phi_l}\,,
\eql{M1bar-mol}\\
\bar q^{ab}_{nl}&=
-e\bra{\phi_n}
\big[r_a-\big(\bar r^a_n+\bar r^a_l\big)/2\big] 
\big[r_b-\big(\bar r^b_n+\bar r^b_l\big)/2\big]
\ket{\phi_l}\,.
\eql{E2bar-mol}
\end{align}
\eql{E1bar-M1bar-E2bar-mol}
\end{subequations}
As they are defined relative to an intrinsic origin located halfway
between the centers $\bar\rr_n=\me{\phi_n}{\rr}{\phi_n}$ and
$\bar\rr_l=\me{\phi_l}{\rr}{\phi_l}$ of the two orbitals, the matrices
$\bar d$, $\bar m$, and $\bar q$ are manifestly origin independent.
For crystals, we find that they acquire ``quantum-geometric'' forms
when expressed in terms of the cell-periodic Bloch eigenstates:
$\bar d$ is given by the interband Berry connection, while $\bar m$
and $\bar q$ are described by generalizations --~with both intraband
and interband parts~-- of the intrinsic orbital
moment~\cite{xiao-rmp10} and of the quantum
metric~\cite{provost-cmp80}, respectively.

The intraband orbital moment and quantum metric were already known to
contribute to the spatially-dispersive optical response in
metals~\cite{zhong-prl16,ma-prb15,gao-prl19}; we clarify that the
quantum metric appears quite generally and not just in two-band
models, and identify additional Fermi-surface terms.  As for the
interband counterparts of the intrinsic orbital moment and quantum
metric, they had not been clearly identified in previous theoretical
studies of spatial dispersion in band
insulators~\cite{natori-jpsj75,zhong-prb93,jonsson-prl96,malashevich-prb10},
where the optical matrix elements were written in velocity form, and
without isolating the magnetic and quadrupolar parts.

The paper is organized as follows. In \sref{basics}, we introduce some
basic definitions and relations. \Sref{derivation} contains the
derivation and analysis of our main result: an expression for the bulk
optical conductivity at first order in $\q$.  The derivation is split
into several steps. We start in \sref{kubo} from the Kubo formula for
the frequency- and wavevector-dependent optical conductivity in the
velocity gauge.  That formula suffers from apparent divergences at
zero frequency, and we recast it in a form that is manifestly
divergence free.  We then expand the Kubo formula to linear order in
$\q$, treating the $\q$ dependence coming from the band dispersion and
from the matrix elements in \srefs{expansion-E}{expansion-M},
respectively.  In \sref{v2r}, we convert the optical matrix elements
from velocity to length form, and express them in terms of intrinsic
multipole transition moments between Bloch eigenstates.  In
\sref{sigma-abc} we collect terms, and report the final expression for
the optical conductivity at first order in $\q$. Finally, in
\sref{magnetoelectric-quadrupolar} we briefly discuss its
decomposition into magnetoelectric and quadrupolar parts, and the
associated physical effects in the static limit. In
\sref{MolecularLimit} we consider the molecular limit of our
formalism, and its relation to the standard multipole theory. The
formal part of the paper ends in \sref{SumRules} with an analysis of
optical sum rules, and in \sref{numerical-results} we present
numerical results for a tight-binding model. We conclude in
\sref{summary} with a summary and discussion.

\section{Basic definitions and relations}
\secl{basics}

Consider the response of a medium to a monochromatic electromagnetic
field.  We work in the ``temporal gauge''~\cite{melrose-book91}, where
the electromagnetic field is fully described by the vector potential
$\A(t,\rr)=\Real\left[ \A(\w,\q)e^{i\left(\q\cdot\rr-\w t\right)}
\right]$. To linear order in the field amplitude, the induced current
density reads
\beq
j_a(\w,\q)=\Pi_{ab}(\w,\q)A_b(\w,\q)=
\frac{1}{i\w}\Pi_{ab}(\w,\q)E_b(\w,\q)\,,
\eql{j-sigma-E}
\eeq
so that one may define an effective conductivity as
$\sigma_{ab}(\w,\q)=(1/i\w)\Pi_{ab}(\w,\q)$~\cite{landau-book84,hornreich-pr68,melrose-book91}.
If the spatial dispersion is weak, the effective conductivity can be
expanded as
\beq
\sigma_{ab}(\omega,\mathbf{q})=
\sigma_{ab}(\omega,\0)+
\sigma_{ab,c}(\omega)q_{c}+
\mathcal{O}(q^2) \ ,
\eql{sigma-expansion}
\eeq
where a summation over the repeated Cartesian index $c$ is implied.
The zeroth-order term is the optical conductivity in the
long-wavelength limit, that is, in the electric-dipole
approximation. The next term in the expansion captures the effects of
spatial dispersion to the order of magnetic dipoles and electric
quadrupoles, and will be the focus of our study.  Let us split it into
symmetric (S) vs antisymmetric (A), and into Hermitian (H) vs
anti-Hermitian (AH) parts with respect to its first two indices,
\begin{subequations}
\eql{sigma-H-AH-S-A}
\begin{align}
\sigma_{ab,c}^{\mathrm{S}}&=
\Real\,\sigma_{ab,c}^{\mathrm{H}} + i\,\Imag\,\sigma_{ab,c}^{\mathrm{AH}} \ ,
\eql{sigma-S}\\
\sigma_{ab,c}^{\mathrm{A}}&=
\Real\,\sigma_{ab,c}^{\mathrm{AH}} + i\,\Imag\,\sigma_{ab,c}^{\mathrm{H}} \ .
\eql{sigma-A}
\end{align}
\end{subequations}
The H and AH parts are absorptive and reactive,
respectively~\cite{landau-book84,melrose-book91}. Regarding the S and
A parts, they transform differently when time reversal $\T$ is applied
to the material, that is, under reversal of its magnetic order
parameter. According to the Onsager reciprocity
relation~\cite{landau-book84,melrose-book91}, the A part is $\T$ even,
and the S part is $\T$ odd; the former describes natural optical
activity, and the latter describes spatially-dispersive
magneto-optical effects.  The entire $\sigma_{ab,c}$ tensor is odd
under spatial inversion $\P$, and hence it vanishes in centrosymmetric
systems.  If both $\P$ and $\T$ are broken but the combined $\P\T$
symmetry is present, $\sigma^\text{A}_{ab,c}$ vanishes but
$\sigma^\text{S}_{ab,c}$ can be nonzero.

\section{The bulk formula for $\sigma_{ab,c}(\w)$}
\secl{derivation}

\subsection{Kubo formula for the effective conductivity
    $\sigma_{ab}(\w,\q)$}
\secl{kubo}

We now specialize to a three-dimensional crystal described by a
single-particle Pauli Hamiltonian $\mathcal{H}$~\cite{blount-ssp62}
with a local external potential, and introduce the electromagnetic
perturbation via an interaction Hamiltonian $\mathcal{H}_{\rm I}$
expressed in the velocity gauge~\cite{rzazewski-jmp04}. To linear
order in the vector potential $\A(t,\rr)$, the interaction Hamiltonian can 
be written as~\cite{zhong-prl16}
\beq
\mathcal{H}_{\rm I}^{(\eta)}(t,\rr)=
\frac{e}{2}\left[
\tilde\A^{(\eta)}(t,\rr)\cdot\v+\v\cdot\tilde\A^{(\eta)}(t,\rr)
\right]+
\frac{e}{m_e}\S\cdot\bpartial_\rr\times\tilde\A^{(\eta)}(t,\rr)\,,
\eeq
where $m_e$ is the electron mass, $\v=(1/i\hbar)[\mathcal{H},\rr]$ is
the unperturbed velocity operator, and $\S$ is the spin operator.  In
addition, we have defined
$\tilde\A^{(\eta)}(t,\rr)=e^{\eta t}\A(t,\rr)$, where the parameter
$\eta$ is formally a positive infinitesimal that controls the
adiabatic turning on of the coupling between the electromagnetic field
and the
crystal~\cite{norman-book18,harrison-book79,dressel2002electrodynamics,bruus2004many,allen06}.

A standard perturbative calculation yields the following Kubo formula
for the optical conductivity in the spectral
representation~\cite{zhong-prl16,dressel2002electrodynamics},
\beq
\sigma^{(\eta)}_{ab}(\w,\q)=
\delta_{ab}\dfrac{ie^{2}N}{m_{e}\w}
+\dfrac{ie^{2}}{\hbar\w}\sum_{n,l}\int_\kk
\dfrac{f_{ln\kk}(\q)}{\w_{ln\kk}(\q)-\w-i\eta}\,
\Mmat_{nl\kk}^{ab}(\q) \ .
\eql{Kubo-VelocityGauge}
\eeq
Here $\int_\kk=\int_{\mathrm{BZ}} d\kk/(2\pi)^{3}$,
$\w_{ln\kk}(\q) = \w_{l}(\kk+\q/2) - \w_{n}(\kk-\q/2)$, where
$\hbar\w_{n}(\kk)=\varepsilon_{n}(\kk)$ is the band energy, and
$f_{ln\kk}(\q) = f_{l}(\kk+\q/2) - f_{n}(\kk-\q/2)$, where
$f_{n}(\kk)=f[\w_{n}(\kk)]$ is the Fermi-Dirac occupation factor.  In
the first term, $N$ is the total number of electrons per unit volume;
in the second, the matrix element is defined as
\beq
\Mmat_{nl\kk}^{ab}(\q)=\left[I^a_{ln\kk}(\q)\right]^*I^b_{ln\kk}(\q) \ ,
\eql{Mnl}
\eeq
where ${\bf I}_{ln\kk}(\q)$ is a sum of orbital and spin
contributions~\cite{zhong-prl16},
\begin{subequations}
\eql{I}
\begin{align}
{\bf I}_{ln\kk}^{\mathrm{orb}}(\q) &=
\me{u_{l}(\kk+\q/2)}{\v(\kk)}{u_{n}(\kk-\q/2)} \ ,
\eql{I-orb}\\
{\bf I}_{ln\kk}^{\mathrm{spin}}(\q) &=\dfrac{ig_{s}}{2m_{e}}
\me{u_{l}(\kk+\q/2)}{\S}{u_{n}(\kk-\q/2)}\times\q \ .
\eql{I-spin}
\end{align}
\end{subequations}
In \eq{I-orb}, $\v(\kk)=(1/\hbar)\bpartial_\kk H(\kk)$ with
$H(\kk)=e^{-i\kk\cdot\rr}\mathcal{H}e^{i\kk\cdot\rr}$, and in
\eq{I-spin}, $g_{s}\approx 2$ is the spin $g$-factor of the electron.
Henceforth, the index $\kk$ will be omitted for brevity.

The $1/\w$ prefactors in \eq{Kubo-VelocityGauge}, inherited from
\eq{j-sigma-E}, make it singular at $\w=0$. That singularity is only
apparent~\cite{allen06}, and it can be removed as follows.  First,
split \eq{Kubo-VelocityGauge} into reactive and absorptive parts using
$\lim_{\eta\rightarrow0^{+}} (x-i\eta)^{-1}=1/x+i\pi\delta(x)$. Next,
notice that the reactive part can be rewritten by invoking the
Kramers-Kr\"onig relation
\beq
\sigma_{ab}^{\mathrm{AH}}(\omega_{0},\q)=
-\dfrac{i}{\pi}\mathrm{P}\int_{-\infty}^{\infty}d\w\,
\dfrac{\sigma_{ab}^{\mathrm{H}}(\omega,\q)}{\omega-\omega_{0}} \ ,
\eql{KK}
\eeq
while in the absorptive part the factor $1/\w$ can be replaced with
$1/\w_{ln}$ thanks to the delta function.  Finally, recombine the two
parts to obtain
\beq
\sigma_{ab}(\ww,\q)=
\dfrac{e^{2}}{\hbar}\sum_{n,l}\int_\kk
\Emat_{ln}(\ww,\q)\Mmat^{ab}_{nl}(\q)\,,
\eql{Kubo-nodiv}
\eeq
where we have defined $\ww=\w+i\eta$ and
\beq
\Emat_{ln}(\ww,\q)=
\dfrac{f_{ln}(\q)}{\w_{ln}(\q)}\dfrac{i}{\w_{ln}(\q)-\ww}\,.
\eql{E-matrix}
\eeq

At zeroth order in $\q$ \eq{Kubo-nodiv} reduces to Eq.~(25) of
Ref.~\cite{allen06}, and at first order it becomes, in the notation of
\eq{sigma-expansion},
\beq
\sigma_{ab,c}(\ww)=\frac{e^2}{\hbar}\sum_{n,l}\int_\kk
\left[
\Emat^{,c}_{ln}(\ww)\Mmat^{ab}_{nl}(\0)+
\Emat_{ln}(\ww,\0) \Mmat^{ab,c}_{nl}
\right]\,.
\eql{sigma-abc-2}
\eeq
In \sref{SumRules}, we discuss how the equivalence between the Kubo
formulas~\eqref{eq:Kubo-VelocityGauge} and~\eqref{eq:Kubo-nodiv} at
zeroth and first orders in $\q$ is related to the oscillator-strength
and rotatory-strength sum rules, respectively. In the context of
tight-binding calculations the diamagnetic term in
\eq{Kubo-VelocityGauge} changes form~\cite{graf-prb95}, while
\eq{Kubo-nodiv} remains unchanged.

In the following subsections, we evaluate the expansion coefficients
of the $\Emat$ and $\Mmat$ matrices appearing in \eq{sigma-abc-2}.  We
start in \sref{expansion-E} with the expansion of $\Emat$, and then
devote \srefs{expansion-M}{v2r} to the expansion and subsequent
manipulations of $\Mmat$, which is where our treatment differs more
substantially from that of previous works.

The terms in the resulting expression for $\sigma_{ab,c}(\tilde\w)$
can be classified as either ``band dispersive'' or ``molecular,''
depending on whether or not they vanish for a crystal composed of
nonoverlapping units. The first term in \eq{sigma-abc-2} is clearly
band dispersive, because the quantity $\Emat^{,c}_{ln}(\tilde\w)$
involves the band velocities $\v_n=\bpartial_\kk\w_n$ (see
\eqs{E1-nn}{E1-ln} below).  While less obvious, the second term in
\eq{sigma-abc-2} is not purely molecular; as we will see, it has a
band-dispersion component that went unnoticed in previous
works~\cite{natori-jpsj75,malashevich-prb10}.

\subsection{Expansion in $\q$ of the band-energy terms}
\secl{expansion-E}

When expanding \eq{E-matrix} in powers of $\q$, the intraband ($l=n$)
and interband ($l\not=n$) parts must be treated separately.  To first
order in $\q$, one finds
\begin{subequations}
\eql{E-expansion}
\begin{align}
\Emat_{nn}(\ww,\0)&=-\dfrac{i}{\ww}f'_n\,,
\eql{E0-nn}\\
\Emat_{ln}(\ww,\0)&=
i\dfrac{f_{ln}}{\w_{ln}}(\w_{ln}+\ww)Z_{ln}(\ww)\,,
\eql{E0-ln}\\
\Emat^{,c}_{nn}(\ww)&=-\dfrac{i}{\ww^2}f'_{n}v_n^c\,,
\eql{E1-nn}\\
\Emat^{,c}_{ln}(\ww)&=\dfrac{i}{2}\dfrac{v_l^c f'_l+v_n^c f'_n}
{\w_{ln}}(\w_{ln}+\ww)Z_{ln}(\ww)\nn
&-if_{ln}\dfrac{Z_{ln}^2(\ww)}{\w_{ln}^2}
\left[
\left(\w_{ln}+\ww\right)^2\left(\w_{ln}-\ww/2\right)
\right]
(v_l^c+v_n^c)\,,
\eql{E1-ln}
\end{align}
\end{subequations}
where $f'_n=\partial f_n/\partial \w_n$, $f_{ln}=f_{l}-f_{n}$,
$\w_{ln}=\w_{l}-\w_{n}$, and 
$Z_{ln}(\tilde{\w})=1/(\w_{ln}^2-\tilde{\w}^2)$.  For the intraband
identities, we used
\beq
\frac{f_{nn}(\q)}{\w_{nn}(\q)}=f_n'+{\cal O}(q^2)\,,
\eql{f-over-w}
\eeq
which follows from $\w_{nn}(\q)= (\partial_a\w_n)q_a+{\cal O}(q^3)$
and $f_{nn}(\q)= (\partial_a f_n)q_a+{\cal O}(q^3)$, where
$\partial_{a}=\partial_{k_{a}}$.

\subsection{Expansion in $\q$ of the optical matrix elements}
\secl{expansion-M}

\subsubsection{Nondegenerate bands}
\secl{nondegen}

Energy eigenstates are only defined up to overall phase factors, and
observable quantities cannot depend on this phase arbitrariness.  In
the case of nondegenerate Bloch bands, physical observables must
remain invariant under single-band quantum-mechanical ``gauge
transformations'' of the form
$\ket{u_{n}} \ \rightarrow \ e^{-i\beta_n}\ket{u_{n}}$, where
$\beta_n$ is a real function of $\kk$.

As the $\Mmat$ matrix defined by \eqs{Mnl}{I} is clearly gauge
invariant, the same must be true for its expansion coefficients
entering \eq{sigma-abc-2} for $\sigma_{ab,c}(\ww)$. When evaluating
those coefficients, we would like to insist that each individual
contribution --~and not just their sum~-- is gauge invariant.  Doing
so will lead to a physically transparent and numerically robust
expression for $\sigma_{ab,c}(\ww)$ in terms of origin-independent
quantities.

The coefficient $\Mmat^{ab}(\0)$ appearing in the first term of
\eq{sigma-abc-2} is trivially gauge invariant, as it involves a single
term,
\beq
\Mmat_{nl}^{ab}(\0)=v^a_{nl}v^b_{ln}\,,
\eql{M0-matrix}
\eeq
which is a product of gauge-covariant velocity matrix elements
(clearly, those matrix elements are also origin independent). Instead,
the coefficient $\Mmat^{ab,c}$ appearing in the second term comprises
several terms, not all of which are individually gauge invariant. The
problematic terms are those that contain matrix elements such as
$\me{u_n}{v_a}{\partial_c u_l}$, because Bloch-state derivatives
transform noncovariantly as
$\ket{\bpartial_\kk u_n} \ \rightarrow \
e^{-i\beta_n}\left(\ket{\bpartial_\kk u_n} - i(\bpartial_\kk\beta_n)
\ket{u_n}\right)$. This can be fixed by writing
$\ket{\bpartial_\kk u_n}$ as $\ket{\D_\kk u_n}-i\AA_n\ket{u_n}$, where
$\ket{\D_\kk u_n}$ is the covariant
derivative~\cite{vanderbilt-book18} and
$\AA_n=\ip{u_n}{i\bpartial_\kk u_n}$ is the intraband Berry
connection. The terms containing Berry connections cancel out, leaving
\begin{align}
\Mmat^{ab,c}_{nl}
&=
\dfrac{1}{2}
\Big(
v_{nl}^a \me{D_c u_l}{v_b}{u_n} +
\me{u_n}{v_a}{D_c u_l}v_{ln}^b - \me{D_c u_n}{v_a}{u_l}v_{ln}^b -
v_{nl}^a\me{u_l}{v_b}{D_c u_n} \Big) \nn
&+ \dfrac{ig_s}{2m_e}
\left(
\epsilon_{acd}S_{nl}^d v_{ln}^b - \epsilon_{bcd}S_{ln}^d v_{nl}^a
\right)\,,
\eql{Dq-M}
\end{align}
where every term is a gauge-invariant product of gauge-covariant
matrix elements, just like in \eq{M0-matrix}.

\subsubsection{Degenerate bands}
\secl{degen}

Gyrotropic birefringence and nonreciprocal directional dichroism occur
in antiferromagnetic crystals such as
Cr$_2$O$_3$~\cite{hornreich-pr68,graham-pmb92}, where the energy bands
are doubly degenerate at every $\kk$ as a result of the combined
$\P\T$ symmetry~\cite{yafet-ssp63}.  To treat such cases, we introduce
degeneracy indices $\lambda$ and $\nu$ for the Bloch states in bands
$l$ and $n$, respectively.  The Kubo formula~\eqref{eq:Kubo-nodiv}
remains unchanged, but the matrix element therein becomes a trace over
the degeneracy indices,
$\Mmat^{ab}_{nl}=\sum_{\lambda,\nu}\,
(I^a_{l\lambda,n\nu})^*I^b_{l\lambda,n\nu}$.  The reasoning leading up
to \eq{Dq-M} follows through, provided that the covariant derivative
is generalized as
$\ket{\D_\kk u_{n\nu}}=\ket{\bpartial_\kk u_{n\nu}} +
i\sum_{\nu'}\,\ket{u_{n\nu'}}\AA_n^{\nu'\nu}$, where
$\AA_n^{\nu'\nu}=\ip{u_{n\nu'}}{i\bpartial_\kk
  u_{n\nu}}$~\cite{vanderbilt-book18}.  The object
$\ket{\D_\kk u_{n\nu}}$ transforms covariantly under multiband gauge
transformations of the form
$\ket{u_{n\nu}}\rightarrow \sum_{\nu'}\,\ket{u_{n\nu'}}U_n^{\nu'\nu}$,
where $U_n$ is a $\kk$-dependent unitary matrix in the degeneracy
indices.  To alleviate the notation, from now on we will assume
nondegenerate bands.

\subsection{Conversion to length (multipole) form}
\secl{v2r}

As we started out from the Kubo formula in the velocity gauge, the
optical matrix elements \eqref{eq:M0-matrix} and \eqref{eq:Dq-M}
entering \eq{sigma-abc-2} for $\sigma_{ab,c}(\ww)$ are written in
terms of the velocity operator.  Now, we would like to recast those
matrix elements in a ``length form'' that brings out their multipole
character. In the case of molecules~\cite{snir-jpc73,pedersen-cpl95},
this is achieved by means of identities such as
$\me{\phi_l}{\v}{\phi_n}= i\w_{ln}\me{\phi_l}{\rr}{\phi_n}$.

In periodic crystals, where the velocity operator is given by the
gradient of the Hamiltonian, the conversion from velocity to length
form follows from the identity
\beq
(\bpartial_\kk H)\ket{u_n} =
(\bpartial_\kk\varepsilon_n)\ket{u_n} - (H-\varepsilon_n)\ket{\D_\kk u_n} \ ,
\eql{v2r-xtal}
\eeq
which can be obtained by differentiating $H\ket{u_n}=\en\ket{u_n}$,
and then writing $\ket{\bpartial_\kk u_n}$ as
$\ket{\D_\kk u_n}-i\AA_n\ket{u_n}$.  Contracting with $\bra{u_l}$
gives $\v_{ln}=\delta_{ln}\v_n+i\w_{ln}\boldsymbol{\A}_{ln}$, where
$\v_n$ is the band velocity, and
\beq
\A_{ln}=
\ip{u_l}{i\mathbf{D} u_n}=(1-\delta_{ln})
\ip{u_l}{i\bpartial_\kk u_n}=
\begin{cases}
\frac{\v_{ln}}{i\w_{ln}}&\text{if $l\not=n$}\,,\\
0&\text{if $l=n$}
\end{cases}
\eql{A-cov}
\eeq
is the interband Berry connection.

With \eqs{v2r-xtal}{A-cov}, one can split \eq{Dq-M} for
$\Mmat^{ab,c}_{nl}$ as
\beq
\Mmat^{ab,c}_{nl}=
i\left( v_{nl}^{a}B_{ln}^{bc} - v_{ln}^{b}B_{nl}^{ac} \right)
+\dfrac{\omega_{ln}}{2}
\left[
(v_{n}^{a}+v_{l}^{a})A^b_{ln} A^c_{nl} +
(v_{n}^{b}+v_{l}^{b})A^a_{nl} A^c_{ln}
\right] \ ,
\eql{M1-matrix}
\eeq
where we have defined
\beq
B_{ln}^{bc} = \dfrac{1}{2i\hbar}
\left(
\bra{D_b u_l}H-\varepsilon_l\ket{D_c u_n} -
\bra{D_c u_l}H-\varepsilon_n\ket{D_b u_n}
\right) 
-\dfrac{g_{s}}{2m_{e}}\epsilon_{bcd}S_{ln}^{d} \ .
\eql{B-matrix}
\eeq
The first term in \eq{M1-matrix} is molecular for $l\not=n$ and band
dispersive for $l=n$, whereas the second term is purely band
dispersive and vanishes for $l=n$ (the distinction between molecular
and band-dispersive contributions was introduced in \sref{kubo}).

From the gauge-covariant and Hermitian matrices $A^a$ and $B^{bc}$, we
can now define for crystals the intrinsic multipole transition moments
that were introduced in \eq{E1bar-M1bar-E2bar-mol} for molecules.  The
intrinsic electric-dipole matrix is $\bar d^a=-e A^a$, while the
intrinsic magnetic-dipole and electric-quadrupole matrices are related
to the antisymmetric and symmetric parts of $B^{bc}$ as follows,
\begin{subequations}
\begin{align}
\bar m^a_{ln}&=\frac{e}{2}\epsilon_{abc}B^{bc}_{ln}\,,
\eql{M1bar-xtal}\\
\bar q^{bc}_{ln}&=\frac{ie}{\w_{ln}}
\left( B^{bc}_{ln}+B^{cb}_{ln}\right)\,.
\eql{E2bar-xtal}
\end{align}
\eql{M1bar-E2bar-xtal}
\end{subequations}
Thus,
\begin{subequations}
\eql{geom}
\begin{align}
\bar\d_{nl}&=-e\ip{u_n}{i\D_\kk u_l}\,,
\eql{E1-geom}\\
\bar\m_{ln}&=\frac{e}{2i\hbar}
\bra{\D_\kk u_l}\times
\left( H-\frac{\el+\en}{2}\right)
\ket{\D_\kk u_n}-
\dfrac{eg_s}{2m_e}\S_{ln}\,,
\eql{M1-geom}\\
\bar q^{bc}_{ln}&=-\dfrac{e}{2}
\left(
\ip{D_b u_l}{D_c u_n} + \ip{D_c u_l}{D_b u_n}
\right)\,,
\eql{E2-geom}
\end{align}
\end{subequations}
where $\bar\m_{ln}$ comprises orbital and spin contributions.  By
expanding the covariant derivatives and then setting $l=n$, one finds
that $\bar\d_{nn}=\0$ [see \eq{A-cov}], and that $\bar\m_{nn}$ and
$-\bar q^{bc}_{nn}/e$ are respectively the intrinsic magnetic moment
$\m_n$~\cite{xiao-rmp10} and the quantum metric
$g^{bc}_n$~\cite{provost-cmp80} of a Bloch eigenstate.

Using \eq{A-cov} for $\A_{ln}$ together with the completeness
relation, \eq{geom} can be recast in a more convenient form for
numerical work,
\begin{subequations}
\eql{sos}
\begin{align}
\bar\d_{nl}&=
\begin{cases}
ie\frac{\v_{nl}}{\w_{nl}}& \text{if $l\not=n$}\\
0                     &  \text{if $l=n$}
\end{cases}\,,
\eql{E1-sos}\\
\bar\m^\text{orb}_{ln}&=\frac{e}{4i}\sum_{p\not= l,n}
\left(\frac{1}{\w_{pl}}+\frac{1}{\w_{pn}}\right)
\v_{lp}\times\v_{pn}\,,
\eql{M1-sos}\\
\bar q^{bc}_{ln}&=-\frac{e}{2}\sum_{p\not= l,n}
\left[
\frac{v^b_{lp}v^c_{pn}}{\w_{pl}\w_{pn}}+(b\leftrightarrow c)
\right]\,.
\eql{E2-sos}
\end{align}
\end{subequations}
As they are written in terms of matrix elements of the velocity
operator, these expressions are manifestly origin independent.  The
correspondence with the molecular expressions in
\eq{E1bar-M1bar-E2bar-mol} will be established in
\sref{MolecularLimit}.

In the case of degenerate bands, \eq{geom} gets modified in the manner
described in \sref{degen}. The modified \eq{sos} remains nonsingular,
as its energy denominator only contains energy differences between
nondegenerate bands.

There is at present considerable interest in quantum-geometric
quantities associated with interband optical
responses~\cite{ma-natmater21}. In this regard, we note that the
quantity $-\bar q^{bc}_{ln}/e$ is distinct from the ``band-resolved
quantum metric''
$g^{bc}_{ln}=\left(A^b_{ln}A^c_{nl}+ A^c_{ln}A^b_{nl}\right)/2$ that
has been introduced in connection with nonlinear optical
responses~\cite{gao-prl20,watanabe-prx21,ahn-natphys22}.  The quantity
$g^{bc}_{ln}$ is gauge invariant for every $l$ and $n$, and when
summed over $l$ it gives the quantum metric of band $n$,
$g^{bc}_n=\Real\,\ip{\partial_b u_n}{\partial_c u_n}-{\cal A}^b_n{\cal
  A}^c_n$. Instead, $-\bar q^{bc}_{ln}/e$ is gauge covariant for
$l\not= n$, and for $l=n$ it reduces to $g^{bc}_n$.

\subsection{Final expression}
\secl{sigma-abc}

We have now gathered all the needed ingredients to evaluate
\eq{sigma-abc-2} for $\sigma_{ab,c}(\ww)$, namely the expansion
coefficients of $\Emat$ in \eq{E-expansion}, and those of $\Mmat$ in
\eqs{M0-matrix}{M1-matrix}. In \eqs{kubo-A}{kubo-S} below, we break
down the resulting expression into antisymmetric ($\T$-even) and
symmetric ($\T$-odd) parts.  The real and imaginary parts of those two
equations are either absorptive or reactive, as per
\eq{sigma-H-AH-S-A}.

To arrive at \eqs{kubo-A}{kubo-S}, several terms containing double
band summations were eliminated by exchanging the $l$ and $n$ indices
(note also that the $l=n$ terms therein vanish, because
$f_{nn}=\w_{nn}=A^a_{nn}=0$).  Those equations are written in terms of
the $A^a$ and $B^{bc}$ matrices, which in turn are related to the
intrinsic multipole transition moments by
\beq
A^a_{nl}=-\frac{1}{e}\bar d^a_{nl}\,,\quad
B^{bc}_{ln}=\frac{1}{e}\bar m^a_{ln}\epsilon_{abc} +
\frac{\w_{ln}}{2ie}\bar q^{bc}_{ln}\,.
\eql{A-B}
\eeq

According to \eq{sos}, the $A^a$ and $B^{ab}$ matrices depend
exclusively on band energies and interband velocity matrix
elements. From the restrictions imposed on these quantities by the
presence of $\P$, $\T$, or $\P\T$ symmetry~\cite{yafet-ssp63}, the
restrictions on $A^a$ and $B^{ab}$ can be deduced. In this way, it may
be verified that the expressions given below satisfy the symmetry
constraints discussed at the end of \sref{basics}.

\subsubsection{Antisymmetric (time-even) part}

The antisymmetric part of $\sigma_{ab,c}(\ww)$ takes the form
\begin{align}
\dfrac{\hbar}{e^{2}}\,\sigma_{ab,c}^{\text{A}}(\tilde{\w})&=
\tilde{\w}\sum_{n,l}\int_\kk Z_{ln}(\ww) 
\bigg\{ 
-f_{ln} \Imag
\left[ A^a_{nl}B^{bc}_{ln}-\flipab \right]\nn
&+
f_{ln}
\left[
\dfrac{1}{2}\left(v^a_n+v^a_l\right)
\Imag\left( A^b_{nl}A^c_{ln}\right)-\flipab
\right]\nn
&+
f_{ln}\left(3\w_{ln}^{2}-\tilde{\w}^{2}\right)
Z_{ln}(\ww)
\Imag\left( A^{a}_{nl}A^{b}_{ln}\right)
\dfrac{1}{2}\left(v^{c}_{n}+v^{c}_{l}\right)\nn
&-
f'_n\w_{ln}
\Imag\left( A^a_{nl}A^b_{ln}\right)v^c_n
\bigg\}\nn
&+\frac{1}{\tilde{\w}}\sum_n \int_\kk f'_n
\left(v^a_n B^{bc}_{nn}-v^b_n B^{ac}_{nn}\right)\,.
\eql{kubo-A}
\end{align}
The five terms in this expression can be classified as follows.  The
first is molecular, while the others are band dispersive; the four
inside curly brackets are interband, while the fifth is intraband; and
the first three are Fermi-sea-like, while the last two are
Fermi-surface-like.

In insulators and cold semiconductors, only the Fermi-sea terms
survive, and one can compare with previous treatments of optical
activity in nonconducting crystals. In
Refs.~\cite{natori-jpsj75,malashevich-prb10}, the sole band-dispersion
contribution to $\sigma_{ab,c}^\text{A}(\ww)$ came from
differentiating the $\Emat$ matrix, that is, from the first term of
\eq{sigma-abc-2} for $\sigma_{ab,c}(\ww)$. It went unnoticed in those
works that the other term in that equation --~where one differentiates
the $\Mmat$ matrix instead~-- is not purely molecular, as shown in
\eq{M1-matrix}.  This is why we have not two but three Fermi-sea terms
in \eq{kubo-A}, one molecular and two band dispersive.

In conductors, the Fermi-surface terms contribute as well. Using
\eq{A-B}, the last term in \eq{kubo-A} becomes
$\left(\epsilon_{acd}K_{bd}-\epsilon_{bcd}K_{ad}\right)/(e\ww)$, where
\beq
K_{ab}=-\sum_n\int_\kk f'_nv^a_n m^b_n
=\sum_n\int_\kk f_n \partial_a m^b_n\,.
\eql{K-ab}
\eeq
This intraband contribution to optical activity involving the
intrinsic magnetic moment of conduction electrons was identified in
Refs.~\cite{zhong-prl16,ma-prb15}, and was evaluated for $p$-doped
tellurium in Refs.~\cite{tsirkin-prb18,sahin-prb18}. The fourth term
in \eq{kubo-A} gives an additional interband contribution to the
optical activity of conductors that was overlooked in previous works.

The low-frequency behavior of the optical rotatory dispersion is
different in insulators and in conductors. For simplicity, let us
consider the propagation of light along the optical axis $z$ of a
uniaxial crystal. The rotatory power is given by~\cite{landau-book84}
\beq
\rho(\w,\tau)=\frac{\w}{2c^2\epsilon_0}
\Real\left[\sigma^\text{A}_{xy,z}(\w+i\tau^{-1})\right]\,,
\eql{rho}
\eeq
where $\epsilon_0$ is the vacuum permittivity and $c$ is the speed of
light. To deal with absorption, the positive infinitesimal $\eta$ in
$\ww=\w+i\eta$ has been reinterpreted heuristically as a
phenomenological scattering rate
$\tau^{-1}$~\cite{barron-book04,allen06,passos-prb18}.  For
frequencies and scattering rates well below the threshold for
interband transitions, $\w,\tau^{-1}\ll\w_{\text{gap}}$, \eq{kubo-A}
yields
\beq
\rho(\w,\tau)=\dfrac{(\w\tau)^{2}}{1+(\w\tau)^{2}}a+b\w^2\,.
\eql{ord}
\eeq
The coefficient $b$ comes from the interband terms which have $\ww$
prefactors, and $a=-(e/c^2\epsilon_0\hbar)K_{xx}$ (with
$K_{xx}=K_{yy}$) comes from the intraband term with a $1/\ww$
prefactor.  In insulators the coefficient $a$ vanishes, and hence the
rotatory power displays the familiar $\w^2$ dependence at low
frequencies~\cite{barron-book04}; in conductors that coefficient is
nonzero, and one can distinguish two different regimes as follows,
\beq
\rho(\w,\tau) \simeq
\begin{cases}
(\tau^{2}a+b)\omega^{2} &\text{if $\w\tau\ll 1$} \ , \\
a+b\w^{2}  &\text{if $\w\tau\gg 1$} \ .
\end{cases}
\eql{ord-metals}
\eeq
In \sref{numerical-results}, we will illustrate these low-frequency
profiles for a concrete tight-binding model.

\subsubsection{Symmetric (time-odd) part}

The symmetric part of $\sigma_{ab,c}(\ww)$ reads
\begin{align}
\frac{\hbar}{e^2}\,\sigma_{ab,c}^\text{S}(\tilde{\w})&=
i\sum_{n,l}\int_\kk Z_{ln}(\ww)
\bigg\{
f_{ln}\w_{ln} \Real
\left[ A^a_{nl}B^{bc}_{ln}+\flipab \right]\nn
&+
f_{ln}\w_{ln}
\left[
\frac{1}{2}\left(v^a_n+v^a_l\right)
\Real\left( A^b_{nl}A^c_{ln}\right)+\flipab
\right]\nn
&-
f_{ln}\w^3_{ln} Z_{ln}(\ww)
\Real\left( A^a_{nl}A^b_{ln}\right)
\left(v^c_n+v^c_l\right)\nn
&+
f'_n\w^2_{ln}
\Real\left( A^a_{nl}A^b_{ln}\right)v^c_n
\bigg\}\nn
&-\frac{i}{\tilde{\w}^2}\sum_n \int_\kk f'_n
v^a_n v^b_n v^c_n\,.
\eql{kubo-S}
\end{align}
The first three terms are Fermi-sea-like, and can be compared with the
expressions obtained for insulators in Ref.~\cite{malashevich-prb10}.
The third is band dispersive, and it corresponds to Eq.~(31) in that
work, while the first two add up to Eq.~(30) therein, revealing its
mixed molecular/dispersive character.

The remaining two terms in \eq{kubo-S} are Fermi-surface-like, and
they can be compared with the expressions obtained for metals in
Ref.~\cite{gao-prl19}. The last term was identified in that work.
Writing $\w^2_{ln}Z_{ln}(\ww)$ as $1+\tilde{\w}^2Z_{ln}(\ww)$ and
noting that $\Real\sum_l A^a_{nl}A^b_{ln}$ is the quantum metric
$g^{ab}_n=-\bar q^{ab}_{nn}/e$ [this can be seen from
\eqs{A-cov}{E2-sos}], the fourth term in \eq{kubo-S} splits into
intraband and interband parts as follows,
\beq
i\sum_n\int_\kk f'_ng^{ab}_n v^c_n+
i\ww^2\sum_{n,l}\int_\kk f'_nZ_{ln}(\ww)
\Real\left(A^a_{nl}A^b_{ln}\right)v^c_n\,.
\eql{g-ident}
\eeq
The intraband piece is similar to the quantity $K_{ab}$ in \eq{K-ab},
but with the intrinsic magnetic moment replaced by the quantum metric
(intrinsic quadrupole moment).  An equivalent result was obtained in
Ref.~\cite{gao-prl19} for two-band models, but without invoking the
identities leading up to \eq{g-ident}, which is what allowed us to
isolate a quantum-metric contribution in the general multiband case.

\subsection{Magnetoelectric and quadrupolar  responses}
\secl{magnetoelectric-quadrupolar}

As recognized in Ref.~\cite{hornreich-pr68}, the physical basis for
$\sigma_{ab,c}(\w)$ is provided by quadrupolar and magnetoelectric
couplings. The quadrupolar couplings are described by a $\T$-odd
totally-symmetric tensor $\gamma_{abc}(\w)$, and the magnetoelectric
ones by a $\T$-odd traceless tensor $\talpha_{ab}(\w)$ together with a
$\T$-even tensor $\calpha_{ab}(\w)$.  Those tensors are defined
as~\cite{hornreich-pr68,malashevich-prb10}
\begin{subequations}
\eql{talpha-calpha-gamma}
\begin{align}
\gamma_{abc}(\w)&=\frac{1}{3i}
\left[
\sigma^\text{S}_{ab,c}(\w)+
\sigma^\text{S}_{bc,a}(\w)+
\sigma^\text{S}_{ca,b}(\w)\right]\,,
\eql{gamma}\\
\talpha_{ab}(\w)&=\frac{1}{3i}\sigma^\text{S}_{ac,d}(\w)\epsilon_{cdb}\,,
\eql{talpha}\\
\calpha_{ab}(\w)&=\frac{1}{4i}\epsilon_{bcd}
\left[\sigma^\text{A}_{cd,a}(\w)-2\sigma^\text{A}_{ac,d}(\w)\right]\,,
\eql{calpha}
\end{align}
\end{subequations}
so that
\begin{subequations}
\eql{repackage}
\begin{align}
\sigma^\text{S}_{ab,c}(\w)&=
i\left[\epsilon_{acd}\talpha_{bd}(\w)+\epsilon_{bcd}\talpha_{ad}(\w)\right]
+i\gamma_{abc}(\w)\,,
\eql{sigma-S-repackage}\\
\sigma^\text{A}_{ab,c}(\w)&=
i\left[\epsilon_{acd}\calpha_{bd}(\w)-\epsilon_{bcd}\calpha_{ad}(\w)\right]\,.
\eql{sigma-A-repackage}
\end{align}
\end{subequations}
Inserting \eq{repackage} in \eq{j-sigma-E} for the induced current
density and then using the Maxwell-Faraday equation yields the
constitutive relation
\beq
j^{\w,\q}_a=
\left(\bnabla\times\M^{\w,\q}\right)_a+
\left[\talpha_{ab}(\w)-\calpha_{ab}(\w)\right]\partial_t B^{\w,\q}_b+
\gamma_{abc}(\w)\nabla_b E^{\w,\q}_c\,,
\eql{j-tot}
\eeq
where ${\bf X}^{\w,\q}$ denotes
${\bf X}(\w,\q)e^{i(\q\cdot\rr-i\w t)}$, and
$M_a^{\w,\q}=\left[\talpha_{ba}(\w)+\calpha_{ba}(\w)\right]E^{\w,\q}_b$.

In the quasi-static limit the electric and magnetic fields become
decoupled, and the three terms in \eq{j-tot} describe separate
physical responses.  The first corresponds to a magnetization current
induced by a uniform electric field; the second to a current induced
by a time-varying magnetic field; and the third to a current induced
by a spatially-varying electric field.  The first two are direct and
inverse magnetoelectric effects~\cite{landau-book84}, and the third is
an electric quadrupolar effect.

The $\T$-even magnetoelectric tensor $\calpha_{ab}$ describes a
``kinetic'' magnetoelectric effect in gyrotropic
conductors~\cite{levitov-jetp85}, while the $\T$-odd tensor
$\talpha_{ab}$ describes magnetoelectric effects in both insulators
and conductors. In the case of insulators, the inverse magnetoelectric
response can be expressed as a polarization current,
$\partial_t P_a^{\w,\q}=\talpha_{ab}(\w)\partial_t B_b^{\w,\q}$;
integrating the adiabatic current in the quasi-static limit, one
obtains the familiar form $P_a=\talpha_{ab}B_b$ of the inverse
magnetoelectric effect~\cite{landau-book84}.  The full $\T$-odd
magnetoelectric tensor includes an additional trace (``axion'') piece
$(\theta e^2/2\pi h)\delta_{ab}$; the axion angle $\theta$ is boundary
sensitive~\cite{vanderbilt-book18}, and this is why it is not captured
by the present formalism, which is based on the electromagnetic
response of a bulk medium.

Quantum-mechanical expressions for the static magnetoelectric and
quadrupolar susceptibilities can be obtained by inserting in
\eq{talpha-calpha-gamma} the formulas given in \sref{sigma-abc} for
$\sigma^\text{S}_{ab,c}(\w)$ and $\sigma^\text{A}_{ab,c}(\w)$,
evaluated at $\w=0$.  This has been done in previous works for
selected terms only: in Ref.~\cite{malashevich-prb10}, the Fermi-sea
orbital contribution to $\talpha_{ab}(0)$ was shown to give the
traceless part of the orbital magnetoelectric susceptibility tensor of
insulators~\cite{malashevich-njp10}; and in Ref.~\cite{zhong-prl16},
the intraband contribution to $\calpha_{ab}(0)$ was shown to reproduce
the kinetic magnetoelectric susceptibility obtained from Boltzmann
transport in the relaxation time approximation combined with the
modern theory of orbital magnetization~\cite{yoda-scirep15}.

Compared to those previous works, the present formalism provides a
more complete description. In particular, it captures $\T$-odd
magnetoelectric and electric-quadrupolar effects in conductors that so
far have only been treated
semiclassically~\cite{gao-prl19,lapa-prb19,xiao-prb21,xiao-prb21b},
and which were found to involve the quantum metric.  A detailed
account of magnetoelectric and electric-quadrupolar responses on the
basis of the present formalism will be given in a separate work.

\section{Molecular limit}
\secl{MolecularLimit}

In this section, we analyze the molecular limit of our formalism.
First, we show how in that limit the bulk expressions for the
intrinsic transition moments $\bar d$, $\bar m$, and $\bar q$ reduce
to those in \eq{E1bar-M1bar-E2bar-mol}. We then show how the formulas
for $\sigma^\text{A}_{ab,c}(\ww)$ and $\sigma^\text{S}_{ab,c}(\ww)$
reduce to the standard molecular expressions in terms of the ordinary
transition moments $d$, $m$, and $q$ in \eq{E1-M1-E2}.

Consider an idealized molecular crystal composed of nonoverlapping
units. For such a crystal, the cell-periodic Bloch states assume the
form~\cite{kohn-pr64,weinreich-book65}
\beq
u_{n\kk}(\rr) \doteq e^{-i\kk\cdot\bxi(\rr)}\phi_n[\bxi(\rr)] \ ,
\eql{bloch-mol}
\eeq
where $\bxi(\rr) = \rr-\R(\rr)$ is the intracell coordinate, $\R(\rr)$
is the lattice vector that folds the absolute coordinate $\rr$ into
the home unit cell, $\phi_n(\rr)$ is vanishingly small outside that
cell, and $\doteq$ denotes an equality that only holds in the
molecular limit.  The intraband Berry connection can now be easily
evaluated by integrating over the home cell,
\beq
\AA_n \doteq \int_\text{cell} d\rr \
\phi_n^*(\rr) e^{i\kk\cdot\rr} \ i\bpartial_\kk
\left[ e^{-i\kk\cdot\rr} \phi_n(\rr) \right] =
\bar\rr_n\,,
\eql{A-supercell}
\eeq
and the covariant derivative of a Bloch state reduces to
\beq
\D_\kk u_{n\kk}(\rr) \doteq -i e^{-i\kk\cdot\rr}
\left( \rr-\bar\rr_n \right) \phi_n(\rr)\,,
\eql{CovDer-mol}
\eeq
for $\rr$ in the home cell. Using this identity in \eq{geom} for
$\bar d$, $\bar m$, and $\bar q$, we recover after some manipulations
the expressions in \eq{E1bar-M1bar-E2bar-mol}. In the case of
$\bar m$, it is necessary to invoke the operator identity
$[v_a,r_b] = [v_b,r_a]$ to rewrite $(\rr\times\v-\v\times\rr)/2$ as
$\rr\times\v$.

With these relations in hand, we can address the molecular limit of
\eqs{kubo-A}{kubo-S} for $\sigma^\text{A}_{ab,c}(\ww)$ and
$\sigma^\text{S}_{ab,c}(\ww)$.  Since the energy bands become
dispersionless in that limit, all band-dispersion terms in those
equations vanish, leaving only the first term in each of them; and
since the transition moments also become independent of $\kk$, we can
set $\int_\kk\doteq 1/V_c$ in those terms ($V_c$ is the cell volume)
to find
\begin{subequations}
\eql{kubo-mol-inv}
\begin{align}
V_c\sigma^\text{A}_{ab,c}&\doteq
\bar G'_{ad}\epsilon_{dbc}+\frac{\ww}{2}\bar a_{abc}-\flipab\,,
\eql{kubo-A-mol}\\
iV_c\sigma^\text{S}_{ab,c}&\doteq
-\bar G_{ad}\epsilon_{dbc}+\frac{\ww}{2}\bar a'_{abc}+\flipab\,,
\eql{kubo-S-mol}
\end{align}
\end{subequations}
where we dropped the $\ww$ dependence for brevity, and defined the
(extensive) molecular tensors
\begin{subequations}
\eql{susc-mol-inv}
\begin{align}
\bar{G}_{ab} &= \frac{1}{\hbar}\sum_{n,l}f_{nl}\w_{ln}Z_{ln}  \Real
\left( \bar{d}_{nl}^{a}\bar{m}_{ln}^{b} \right)  \ ,
\eql{G-bar}\\
\bar{G}_{ab}' &= -\frac{1}{\hbar}\sum_{n,l}f_{nl}
	\tilde{\w} Z_{ln}  \Imag
	\left( \bar{d}_{nl}^{a}\bar{m}_{ln}^{b} \right) \ ,
\eql{Gprime-bar}\\
\bar{a}_{abc} &= \frac{1}{\hbar}\sum_{n,l}f_{nl}\omega_{ln}Z_{ln} \Real
\left( \bar{d}_{nl}^{a}\bar{q}_{ln}^{bc}\right) \ ,
\eql{a-bar}\\
\bar{a}_{abc}' &= -\frac{1}{\hbar}\sum_{n,l}f_{nl}
	\dfrac{\w_{ln}^{2}}{\tilde{\w}} Z_{ln} \Imag
\left( \bar{d}_{nl}^{a}\bar{q}_{ln}^{bc} \right)  \ .
\eql{aprime-bar}
\end{align}
\end{subequations}

Writing $f_{nl}$ as $f_{n}(1-f_{l}) - f_{l}(1-f_{n})$, the $f_{nl}$
factors in the expressions above can be replaced with
$2f_n(1-f_l)$. In that form, $\bar{G}$, $\bar{G}'$ and $\bar{a}$
become single-particle versions of the multipolar susceptibility
tensors $G$, $G'$ and $a$ defined in Eqs.~(2.83), (2.85) and (2.86) of
Ref.~\cite{raab-book05}, with one difference: the ordinary transition
moments $d$, $m$, and $q$ have been replaced with $\bar d$, $\bar m$,
and $\bar q$.  It is not immediately clear that the same is true for
$\bar a'$, since \eq{aprime-bar} contains a factor of $\w^2_{ln}/\ww$
in place of the $\ww$ factor appearing in Eq.~(2.84) of
Ref.~\cite{raab-book05} for $a'$. However, those factors are
interchangeable in the expression for $a'$, as can be seen in the
manner described around around Eqs.~(2.75--2.78) of
Ref.~\cite{raab-book05}.  Thus, $(\bar G,\bar G',\bar a,\bar a')$ are
origin-independent versions of the molecular tensors $(G,G',a,a')$
entering the standard multipole theory.
 
Next, let us consider the propagation of light inside our idealized
molecular crystal.  For a given propagation direction $\hat{\bf n}$,
we define (intensive) optical-activity and gyrotropic-birefringence
tensors as
$\beta^\text{A}_{ab}(\hat{\bf n})=-\sigma^\text{A}_{ab,c}\hat n_c$ and
$\beta^\text{S}_{ab}(\hat{\bf n})=i\sigma^\text{S}_{ab,c}\hat n_c$,
respectively.  Using \eq{kubo-mol-inv}, we obtain Eqs.~(5.8) and~(5.9)
of Ref.~\cite{raab-book05} for those tensors, but with
$(\bar G,\bar G',\bar a,\bar a')$ in place of $(G,G',a,a')$.
Inserting \eq{E1bar-M1bar-E2bar-mol} in \eq{susc-mol-inv}, the terms
containing the orbital centers drop out from the combinations of
molecular tensors appearing in \eq{kubo-mol-inv}.  Thus,
$(\bar d,\bar m,\bar q)$ can be safely replaced by $(d,m,q)$ for the
purpose of evaluating the optical properties of an idealized molecular
crystal.  This completes the proof that our formalism correctly
reduces to the standard single-particle multipole theory in the
molecular limit.

In summary, we have in \eqs{kubo-mol-inv}{susc-mol-inv} a
reformulation of the molecular multipole theory of optical spatial
dispersion at linear order in $\q$ in terms of
translationally-invariant property tensors. This is in contrast to the
standard formulation, where translational invariance is achieved by a
cancellation between the origin dependences of the magnetic-dipole and
electric-quadrupole
terms~\cite{barron-book04,raab-book05,norman-book18,pedersen-cpl95}.

\section{Sum rules}
\secl{SumRules}

In \sref{kubo}, we wrote two alternative Kubo formulas for
$\sigma_{ab}(\w,\q)$, namely \eqs{Kubo-VelocityGauge}{Kubo-nodiv}.
The former displays apparent $1/\w$ divergences at $\w=0$, whereas the
latter is explicitly divergence free.  In this section, we scrutinize
the mathematical identities that underlie the equivalence between
those two formulas at zeroth and first order in $\q$, and relate those
identities to the oscillator- and rotatory-strength sum rules.

\subsection{Equivalence between the two forms of the Kubo formula}
\secl{equivalence}

Let us denote as $(ie^{2}/\omega)\Delta_{ab}(\mathbf{q})$ the
difference between the reactive parts of the Kubo
formulas~\eqref{eq:Kubo-VelocityGauge}
and~\eqref{eq:Kubo-nodiv}. Writing $1/[x(a-x)]$ as
$(1/a)[1/x+1/(a-x)]$, we find
\beq
\Delta_{ab}(\q) = \delta_{ab}\dfrac{N}{m_e}+
\dfrac{1}{\hbar}\sum_{n,l}\int_\kk\dfrac{f_{ln}(\q)}{\w_{ln}(\q)}
M_{nl}^{ab}(\q) \ ,
\eql{Delta-def}
\eeq
and using
\begin{subequations}
\eql{q-Identities}
\begin{align}
\w_{ln}(\q)&=-\w_{nl}(-\q)\,,
\eql{w-q}\\
f_{ln}(\q)&=-f_{nl}(-\q)\,,
\eql{f-q}\\
M^{ab}_{nl}(\q)&=\left[ M^{ab}_{ln}(-\q)\right]^*\,,
\eql{M-q}\\
M^{ab}_{nl}(\q)&=\left[ M^{ba}_{nl}(\q)\right]^*\,,
\eql{M-ab-ba}
\end{align}
\end{subequations}
we obtain
\begin{subequations}
\eql{Re-Im-Delta-symm}
\begin{align}
\Real\,\Delta_{ab}(\q)&=\Real\,\Delta_{ba}(\q)=\Real\,\Delta_{ab}(-\q)\,,
\eql{Re-Delta-symm}\\
\Imag\,\Delta_{ab}(\q)&=-\Imag\,\Delta_{ba}(\q)=-\Imag\,\Delta_{ab}(-\q)\,.
\eql{Im-Delta-symm}
\end{align}
\end{subequations}
For the two Kubo formulas to be equivalent, $\Delta_{ab}(\q)$ must
vanish identically, and according to the derivation in \sref{kubo}
this is guaranteed by the Kramers-Kr\"onig relations.  To analyze the
behavior of $\Delta_{ab}(\q)$ at zeroth and first order in $\q$, we
expand it as
\beq
\Delta_{ab}(\q) = \Delta_{ab}(\0) + \Delta_{ab,c}q_c +
\mathcal{O}(q^{2}) \ .
\eql{Delta-expansion}
\eeq

Let us start with the zeroth-order term in the expansion. Writing the
electron density $N$ in the first term of \eq{Delta-def} as
$\sum_n\int_\kk f_n$, and using the identity in \eq{f-over-w},
followed by an integration by parts, to deal with the $l=n$
contribution to the second term, we obtain
\beq
\Delta_{ab}(\0) = \sum_n\int_\kk f_n
\left[
\dfrac{\delta_{ab}}{m_e} + 2\sum_{l\neq n}\dfrac{\Real\left(
    v_{nl}^{a}v_{ln}^{b} \right)} {\varepsilon_n-\varepsilon_l} -
  \dfrac{1}{\hbar^{2}}\dfrac{\partial^{2}\varepsilon_n} {\partial
    k_a\partial k_b}
\right] =0 \ .
\eql{D-0}
\eeq
The quantity in square brackets is formally real and symmetric in
accordance with \eq{Re-Delta-symm}, and it vanishes identically by
virtue of the effective-mass theorem. We note that the same theorem
was invoked in Ref.~\cite{sipe-prb93} to remove the apparent
divergence at $\w=0$ in the dielectric function of insulators and
semiconductors.

In preparation for analyzing the first-order term in
\eq{Delta-expansion}, let us compare \eq{Delta-def} for
$\Delta_{ab}(\q)$ with the $\w\rightarrow 0$ limit of
$\Pi_{ab}(\w,\q)=i\w\sigma_{ab}(\w,\q)$ [see \eq{j-sigma-E}],
evaluated using the Kubo formula~\eqref{eq:Kubo-VelocityGauge}. This
gives $-e^2\Delta_{ab}(\q)=\Pi_{ab}(0,\q)$, and so
$-e^2\Delta_{ab,c}=\left.\partial_{q_c}\Pi_{ab}(0,\q)\right|_{\q=\0}$.
From the analysis of this quantity in Ref.~\cite{zhong-prl16} (see
Sec.~III.A.1 of its Supplemental Material), we conclude that
\beq
\Delta_{ab,c} =
-\dfrac{i}{\hbar}\epsilon_{abc}\sum_n\int_\kk f_n\,
\v_{n}\cdot\boldsymbol{\Omega}_n =0 \ ,
\eql{D-1}
\eeq
where $\boldsymbol{\Omega}_n$ is the Berry curvature.  In agreement
with \eq{Im-Delta-symm}, the expression above is purely imaginary and
antisymmetric in $a$ and $b$. It vanishes identically for topological
reasons, and that amounts to a no-go theorem for the chiral magnetic
effect in equilibrium~\cite{zhong-prl16}.

In a recent work~\cite{chang-prb22}, an expression was derived for
$\sigma_{ab,c}(\w)$ that contains a term diverging as $1/\w$. The
authors found that the prefactor of that term was neglibible for a
specific tight-binding model, but they were unable to confirm
analytically that it vanishes in general. The removal of that apparent
divergence can be achieved by means of \eq{D-1}.

The vanishing of $\Delta_{ab}(\mathbf{q})$ dictates the high-frequency
behavior of the optical conductivity as follows.  Suppose there is a
frequency $\w_\text{max}$ above which the system does not
absorb~\cite{barron-book04}; setting $\w\gg\w_\text{max}$ in
\eqs{Kubo-nodiv}{E-matrix} and comparing with \eq{Delta-def}, we can
deduce that
\beq
\sigma_{ab}(\w\gg\w_{\mathrm{max}},\q) =
\delta_{ab}\dfrac{ie^{2}N}{\w m_{e}}
-\frac{ie^2}{\w}\Delta_{ab}(\q)\, .
\eeq
Thus, at high frequencies the optical conductivity reduces to the
diamagnetic term; and since that term is independent of $\q$, we
conclude that $\sigma_{ab}(\w,\0)$ decays as $1/\w$.

In Sec.~5.2.4 of Ref.~\cite{barron-book04}, the high-frequency
behavior of the optical activity of molecules was inferred from the
rotatory-strength sum rule. This is consistent with the present
analysis, because that sum rule is a direct consequence of the
vanishing of $\Delta_{ab,c}$, as we will now show.

\subsection{Optical sum rules}

Consider the sum rules obtained by integrating over positive
frequencies the absorptive part of the optical conductivity, taking
into account both interband and intraband absorption.  Writing
$\int_0^\infty f(\w)d\w$ as $\langle f(\w)\rangle$ and using
\eq{sigma-H-AH-S-A} yields
\beq
\left< \sigma^\text{H}_{ab}(\w,\q) \right>=
\left< \Real\,\sigma^\text{S}_{ab}(\w,\q) \right>+
i\left< \Imag\,\sigma^\text{A}_{ab}(\w,\q) \right>\,.
\eql{H-S-A}
\eeq
To evaluate this quantity, we begin by taking the Hermitian part of
\eq{Kubo-nodiv} for $\eta\rightarrow 0^+$,
\beq
\sigma^\text{H}_{ab}(\w,\q)=-\frac{\pi e^2}{\hbar}\sum_{n,l}\int_\kk
\frac{f_{ln}(\q)}{\w_{ln}(\q)}\Mmat^{ab}_{nl}(\q)
\delta\left[\w-\w_{ln}(\q)\right]\,.
\eql{sigma-H}
\eeq
Making the substitution
\beq
f_{ln}(\q)=f_l(\kk+\q/2)\left[1-f_n(\kk-\q/2)\right]
-f_n(\kk-\q/2)\left[1-f_l(\kk+\q/2)\right]\,,
\eeq
and noting that at zero temperature only the second term contributes
to \eq{sigma-H} when $\w>0$ and $\q\approx\0$, we obtain
\beq
\left< \Real\,\sigma^\text{S}_{ab}(\w,\q) \right>+
i\left< \Imag\,\sigma^\text{A}_{ab}(\w,\q) \right>
= R_{ab}(\q) + i\, I_{ab}(\q)\,,
\eql{sigma-RI}
\eeq
where we have defined
\begin{subequations}
\begin{align}
R_{ab}(\q) &= \dfrac{\pi e^{2}}{\hbar}\sum_{n,l}\int_\kk f_{n}(\kk-\q/2)
\left[ 1-f_{l}(\kk+\q/2) \right] \dfrac{\Real \left[ \Mmat_{nl}^{ab}(\q)
\right]}{\w_{ln}(\q)} \,, \eql{R-ab} \\
I_{ab}(\q) &= \dfrac{\pi e^{2}}{\hbar}\sum_{n,l}\int_\kk f_{n}(\kk-\q/2)
\left[ 1-f_{l}(\kk+\q/2) \right] \dfrac{\Imag\left[ \Mmat_{nl}^{ab}(\q)
\right]}{\w_{ln}(\q)} \,. \eql{I-ab}
\end{align}
\end{subequations}
Let us split $R_{ab}(\q)$ and $I_{ab}(\q)$ in \eq{sigma-RI} into even
and odd parts in $\q$. Using the identities in \eq{q-Identities}, one
finds that the even part of $R_{ab}(\q)$ plus the odd part of
$iI_{ab}(\q)$ is proportional to the second term in \eq{Delta-def} for
$\Delta_{ab}(\q)$.  Therefore,
\begin{subequations}
\eql{opt-sum-rule}
\begin{align}
\left<\Real\,\sigma^\text{S}_{ab}(\w,\q)\right>&=\frac{\pi e^2}{2} \left[
\delta_{ab}N/m_e
-\Real\,\Delta_{ab}(\q) \right]
+\frac{1}{2}\left[R_{ab}(\q)-R_{ab}(-\q)\right]\,,
\eql{sum-rule-Re-S}\\
\left<\Imag\,\sigma^\text{A}_{ab}(\w,\q)\right>&=
-\frac{\pi e^2}{2}\Imag\,\Delta_{ab}(\q)
+\frac{1}{2}\left[I_{ab}(\q)+I_{ab}(-\q)\right]\,,
\eql{sum-rule-Im-A}
\end{align}
\end{subequations}
where we keep track of the vanishing quantity $\Delta_{ab}(\q)$.

The expansion of \eq{opt-sum-rule} in powers of $\q$ generates a
series of sum rules. Since, according to \eq{Re-Im-Delta-symm}, the
terms $\Real\,\Delta_{ab}$ and $\Imag\,\Delta_{ab}$ only contribute
(formally) at even and odd orders in $\q$, respectively, and since the
reverse is true for the second terms in
\eqs{sum-rule-Re-S}{sum-rule-Im-A}, we obtain
\begin{subequations}
\eql{sum-rules}
\begin{align}
\left<\Real\,\sigma^\text{S}_{ab}(\w,\0)\right>&=\frac{\pi e^2}{2}
\left[
\delta_{ab}N/m_e -\Delta_{ab}(\0) \right]\,,
\eql{f-sum-rule}\\
\left<\Imag\,\sigma^\text{A}_{ab}(\w,\0)\right>&=I_{ab}(\0)\,,
\eql{dichroic-f-sum-rule}\\
\left<\Real\,\sigma^\text{S}_{ab,c}(\w)\right>&=R_{ab,c}\,,
\eql{NDD-sum-rule}\\
\left<\Imag\,\sigma^\text{A}_{ab,c}(\w)\right>&=
-\frac{\pi e^2}{2}\Delta_{ab,c}\,,
\eql{rot-sum-rule}
\end{align}
\end{subequations}
to linear order in $\q$. Below, we consider each of these identities
in turn.

Once we set $\Delta_{ab}(\0)=0$ in accordance with \eq{D-0},
\eq{f-sum-rule} becomes the oscillator-strength sum
rule~\cite{landau-book84}
\beq
\left<\Real\,\sigma^\text{S}_{ab}(\w,\0)\right>=
\frac{\w^2_\text{p}}{8}\delta_{ab}\,,
\eql{f-sum-rule-wp}
\eeq
where $\w_\text{p}=(4\pi e^2N/m_e)^{1/2}$ is the plasma frequency.  As
already mentioned, for tight-binding models the diamagnetic term in
\eq{Kubo-VelocityGauge} changes form while \eq{Kubo-nodiv} remains
unchanged, which leads to a modified oscillator-strength sum
rule~\cite{graf-prb95}.

\Eq{dichroic-f-sum-rule} is the rotatory-strength sum rule for
magnetic circular dichroism.  At $\q=\0$, the intraband part of
\eq{I-ab} vanishes because $\Mmat_{nn}^{ab}(\0)$ is real, and from the
interband part we recover the bulk expression given in
Ref.~\cite{souza-prb08} for that sum rule. If a single band is
occupied, the integrated magnetic circular dichroism spectrum is
proportional to the intrinsic orbital magnetic moment of the Bloch
states in that band~\cite{souza-prb08,yao-prb08}.

\Eq{NDD-sum-rule} is a sum rule for nonreciprocal directional
dichroism. An explicit expression can be obtained by expanding
\eq{R-ab} to first order in $\q$.

Finally, by setting $\Delta_{ab,c}=0$ in \eq{rot-sum-rule} in
accordance with \eq{D-1}, we arrive at the rotatory-strength sum rule
for natural circular dichroism,
\beq
\left<\Imag\,\sigma^\text{A}_{ab,c}(\w)\right>=0\,.
\eeq
This sum rule is well known for molecules in
solution~\cite{condon-rmp37}, as well as for oriented
molecules~\cite{barron-book04}. Here, we have relied on a topological
argument~\cite{zhong-prl16} to show that it remains valid for
crystals, both insulating and conducting. Alternative discussions
restricted to insulators are given in
Refs.~\cite{natori-jpsj75,zhong-prb93}.

The above derivation highlights the connection between the
oscillator-strength and natural rotatory-strength sum rules for
crystals, and the equivalence between the Kubo formulas
\eqref{eq:Kubo-VelocityGauge} and~\eqref{eq:Kubo-nodiv} --~that is,
the vanishing of $\Delta_{ab}(\q)$~-- at zero and first order in $\q$,
respectively.  More generally, the expansion of \eq{opt-sum-rule} in
powers of $\q$ yields at each order two optical sum rules, one of
which relies on the vanishing of $\Delta_{ab}(\q)$ at that order.

\section{A tight-binding example}
\secl{numerical-results}

In this section, we use numerical tight-binding calculations to
validate our formalism, and to illustrate the distinctive
low-frequency profiles of the rotatory power in insulators and
conductors.

\begin{figure}[!t]
\centering
\includegraphics{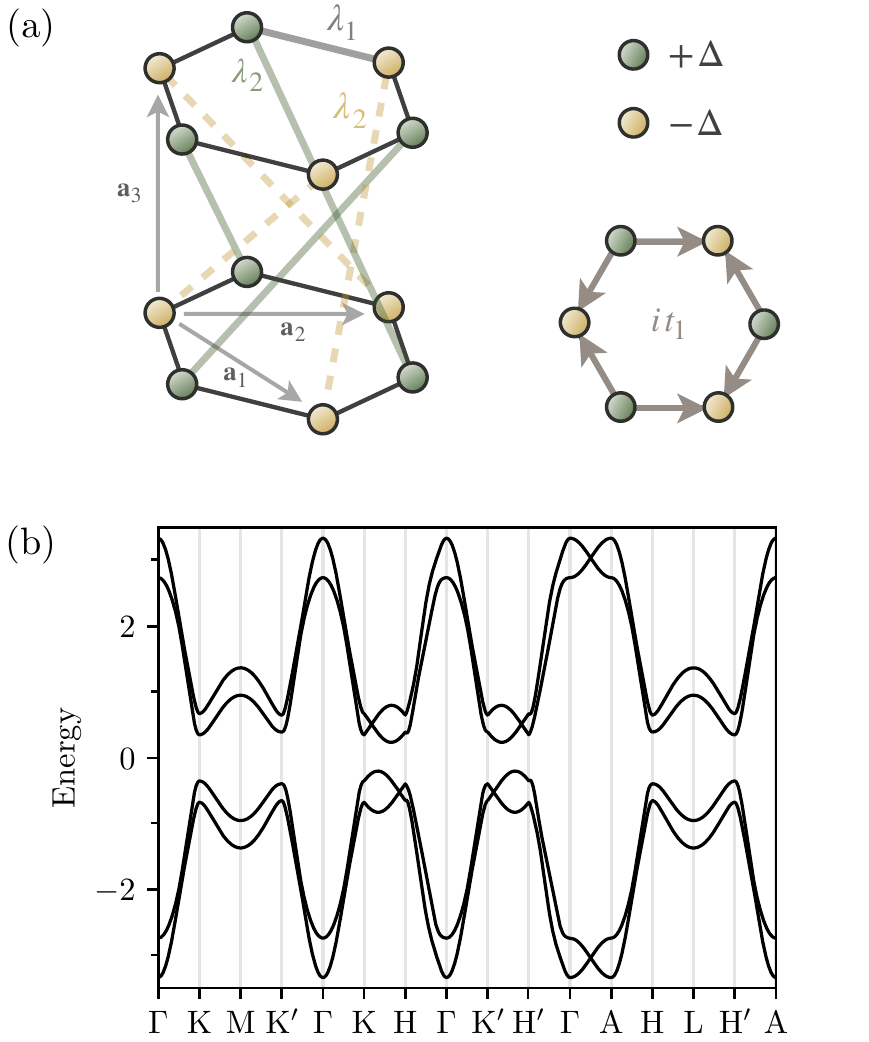}
\caption{(a) The tight-binding model of \eq{TB-model}. The
  crystallographic vectors are $\mathbf{a}_{1}=(\sqrt{3}a,0,0)$,
  $\mathbf{a}_{2}=(\sqrt{3}a/2,3a/2,0)$, and $\mathbf{a}_{3}=(0,0,c)$,
  with $a$ the distance between nearest-neighbor sites on the same
  layer, and $c$ the interlayer distance. The on-site energies and
  hoppings are indicated schematically.  As shown on the right, the
  complex intralayer hoppings from $+\Delta$ sites to $-\Delta$ sites
  have amplitude $it_1$; the reverse hoppings (not shown) have
  amplitude $-it_1$. (b) Band structure of the model for the
  Hamiltonian parameters given in the main text.}
\figl{fig1}
\end{figure}

As a simple model of a bulk crystal with nonzero
$\sigma^\text{A}_{ab,c}$, we take the tight-binding model of
Ref.~\cite{yoda-scirep15}, which consists of honeycomb layers coupled
by a chiral pattern of interlayer hoppings.  To break time-reversal
symmetry, so that $\sigma^\text{S}_{ab,c}$ becomes nonzero as well, we
add complex intralayer hoppings.  The resulting model is depicted in
\fref{fig1}(a), and its Hamiltonian reads
\beq
\mathcal{H} = \Delta\sum_{i}\xi_{i} c_{i}^{\dagger}c_{i} +
it_{1}\sum_{\langle i,j \rangle} \xi_{j} c_{i}^{\dagger}c_{j} +
\dfrac{i\lambda_1}{a}\sum_{\langle i,j \rangle} c_{i}^{\dagger}
\left({\boldsymbol\sigma}\cdot{\boldsymbol\delta}_{ij}\right) c_j
+\dfrac{i\lambda_2}{a}\sum_{[i,j]} c_i^\dagger
\left({\boldsymbol\sigma}\cdot {\bf d}_{ij}\right) c_{j}  \ .
\eql{TB-model}
\eeq
The first term is a staggered on-site potential, with $\xi_i=\pm 1$
for the two sublattices in each layer.  The second and third terms
describe intralayer hoppings between nearest-neighbor sites $i$ and
$j$: the second is the complex hopping responsible for breaking time
reversal, and the third is a spin-orbit coupling term; therein,
${\boldsymbol\sigma}$ is the vector of Pauli matrices and
${\boldsymbol \delta}_{ij}$ is the vector taking from site $j$ to site
$i$. The last term is the helical pattern of interlayer hoppings that
renders the model chiral, with ${\bf d}_{ij}$ the vector taking from
site $j$ to site $i$ in adjacent layers.  We choose the distance $a$
between nearest-neighbor sites on the same layer as the unit of
length, and the nearest-neighbor hopping amplitude $t_1$ as the unit
of energy. For our tests, we set $c=1$, $\Delta=0.5$,
$\lambda_1=-0.06$, and $\lambda_2=0.05$.

Exploiting the translational symmetry of the crystal, we replace the
site indices $\{i\}$ with $\{\R i\}$, where the lattice vector $\R$
labels the cell, and $i$ is now an intracell site index.  The
Hamiltonian matrix elements are denoted by
$\mathcal{H}_{ij}(\R)=\me{\phi_{\0 i}}{\mathcal{H}}{\phi_{\R j}}$,
where $\phi_{\R j}(\rr)=\varphi_j(\rr-\R-\btau_j)$ is a basis orbital
centered at $\R-\btau_j$~\cite{vanderbilt-book18}.  The tight-binding
Hamiltonian in $\kk$ space is constructed as
\beq
H^\kk_{ij} =
\sum_\R e^{i\kk\cdot(\R+\btau_j-\btau_i)}\mathcal{H}_{ij}(\R)\,,
\eql{H-TB}
\eeq
leading to the eigenvalue equation
$H_\kk\cdot C_{n\kk}=\varepsilon_{n\kk}C_{n\kk}$.

\begin{figure}[!t]
\centering
\includegraphics{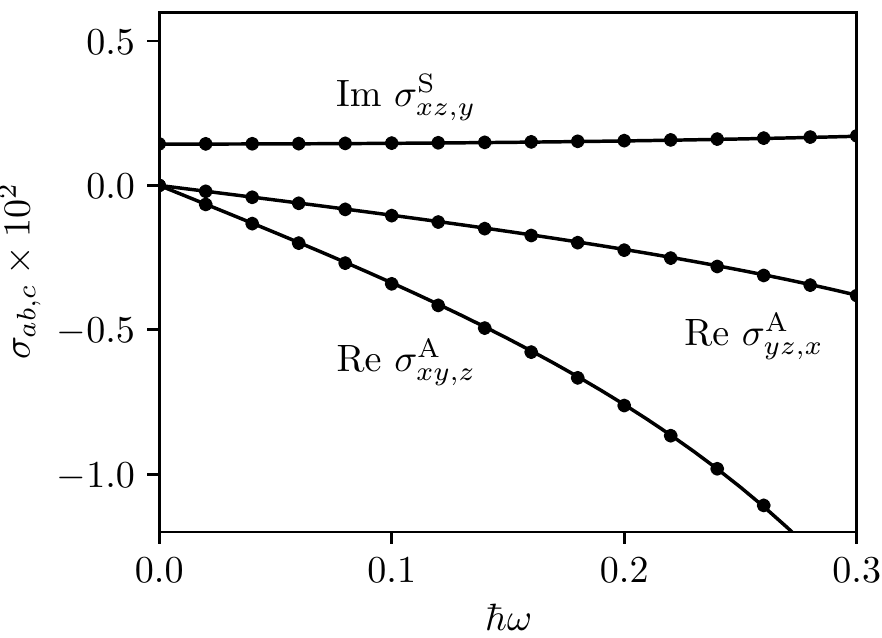}
\caption{ Numerical results for the nonzero components of
  $\sigma_{ab,c}$ for the model of \eq{TB-model} and \fref{fig1}, with
  the two lowest bands treated as occupied. The results are plotted as
  a function of frequency up to $\hbar\w=0.3$, which is well below the
  threshold for interband absorption
  ($\varepsilon_\text{gap}\approx 0.53$). Filled circles:
  extrapolation from finite-size crystallites.  Solid lines: bulk
  crystal. The tensor $\sigma_{ab,c}$ has been divided by $e^2/\hbar$
  to make it dimensionless.}
\figl{fig2}
\end{figure}

The energy bands of the model are displayed in \fref{fig1}(b). There
are two composite groups with two bands each, separated by a gap. We
treat the lowest group as occupied, and calculate
$\sigma^\text{A}_{ab,c}$ and $\sigma^\text{S}_{ab,c}$ at zero
temperature using \eqs{kubo-A}{kubo-S}, respectively.  The $A^a$ and
$B^{ab}$ matrices are evaluated from \eqs{sos}{A-B}, with effective
velocity matrix elements given by~\cite{graf-prb95}
\beq
\v_{nl}(\kk) = \frac{1}{\hbar}C_{n\kk}^{\dagger}\cdot
\left(\bpartial_\kk H_\kk\right)\cdot C_{l\kk} \,.
\eeq
As the system is insulating, the only nonzero contributions to
$\sigma_{ab,c}$ come from the Fermi-sea terms in
\eqs{kubo-A}{kubo-S}; we will restrict our calculations to frequencies
well below the threshold for interband absorption, where one can
safely set $\eta=0$ in those equations.

The magnetic point group of the model is 32.  Of the four independent
tensor components that are allowed by
symmetry~\cite{newnham2005properties,Gallego:lk5043},
$\sigma_{yz,x}^{\mathrm{A}}$, $\sigma_{xy,z}^{\mathrm{A}}$,
$\sigma_{xx,y}^{\mathrm{S}}$ and $\sigma_{xz,y}^{\mathrm{S}}$, only
the first three are actually nonzero when the Fermi level
$\varepsilon_\text{F}$ lies in the gap. Converged results, obtained by
sampling the Brillouin zone on a uniform mesh of
$50\times 50\times 50$ $k$ points, are shown as solid lines in
\fref{fig2}.

For comparison, we show as filled circles in \fref{fig2} the results
obtained from calculations on finite crystallites. We treat them as
``molecules,'' and evaluate $\sigma^\text{A}_{ab,c}$ and
$\sigma^\text{S}_{ab,c}$ from \eq{kubo-mol-inv} under open boundary
conditions. Calculations are performed for samples with $L+1$ unit
cells in each crystallographic direction, with $L$ ranging from 4 to
12. The results are extrapolated to $L\rightarrow\infty$ by fitting
them to the function
\beq
f(L) = f_0+f_1/L+f_2/L^2+f_3/L^3\,,
\eeq
where $f_0$ is the extrapolated value to be compared with the result
of the bulk calculation, and $f_1/L$, $f_2/L^2$, and $f_3/L^3$ account
for face, edge, and corner corrections,
respectively~\cite{ceresoli-prb06}.  The excellent agreement seen in
\fref{fig2} between the two types of calculations confirms the
validity of our formalism for band insulators.

In \fref{fig3}, we take the $\sigma^\text{A}_{xy,z}$ curves from
\fref{fig2} and split them into origin-independent contributions. For
the extrapolated crystallites (left panel), there are two types of
contributions in \eq{kubo-mol-inv}: those containing $\bar m$ are
denoted as $\MDbar$, and those containing $\bar q$ are denoted as
$\EQbar$. For the bulk crystal (right panel), there are, in addition
to $\MDbar$ and $\EQbar$ contributions from the first term in
\eq{kubo-A}, band-dispersion contributions from the second and third
terms, which are denoted as $\v$.

A comparison between the two panels of \fref{fig3} reveals that the
$\MDbar$ and $\EQbar$ contributions are different for the extrapolated
crystallites and for the bulk crystal, with the difference being
exactly compensated by the $\v$ contributions that are only present in
the latter.  A similar situation occurs for the ground-state orbital
magnetization of a crystal, whose bulk expression contains a subtle
Berry-curvature term without an obvious counterpart in the molecular
theory~\cite{thonhauser-prl05,ceresoli-prb06}. For a two-dimensional
insulator with a single valence band, that term reads
$-(e/2\hbar)\int_\kk \varepsilon_\kk(\partial_x {\mathcal
  A}_y-\partial_y {\mathcal A}_x)$ or, after integrating by parts,
$(e/2)\int_\kk (v_x{\mathcal A}_y-v_y{\mathcal A}_x)$. As in
\fref{fig3}, this additional band-dispersion contribution must be
included in the bulk calculation to recover the net orbital
magnetization of a large flake~\cite{thonhauser-prl05,ceresoli-prb06}.

To conclude, let us illustrate the different low-frequency behaviors
of the rotatory power in insulating and conducting states of our
model.  We evaluate $\rho(\w,\tau)$ for the bulk model from
\eqs{kubo-A}{rho}, setting $\hbar/\tau=2\times10^{-3}$. The frequency
range is chosen as $0\leq \hbar\w\leq 10^{-2}$, and the calculation is
carried out at zero temperature for $\varepsilon_\text{F}=0.0$
(insulator) and $\varepsilon_\text{F}=1.0$ (metal). In both cases, a
uniform mesh of $100\times100\times100$ $k$ points is used to sample
the Brillouin zone; to improve the convergence of the calculation in
the metallic case, the Fermi-surface terms in Eq.~\eqref{eq:kubo-A}
are evaluated as Fermi-sea integrals by performing an integration by
parts.

The resulting $\rho(\w)$ profiles, plotted in \fref{fig4} as solid
lines, display the behavior dictated by \eq{ord}. The insulator
displays a simple $\rho\propto\w^2$ decay with a negligible influence
from the scattering time $\tau$. Instead, for the metal one can
distinguish two different parabolic regimes (dashed lines) delimited
by $\w\tau\sim 1$, in accordance with \eq{ord-metals}.  This
distinctive low-frequency profile of the optical rotatory dispersion
in conducting crystals awaits experimental verification.

\begin{figure*}[!t]
	\centering
	\includegraphics[width=1.03\linewidth]{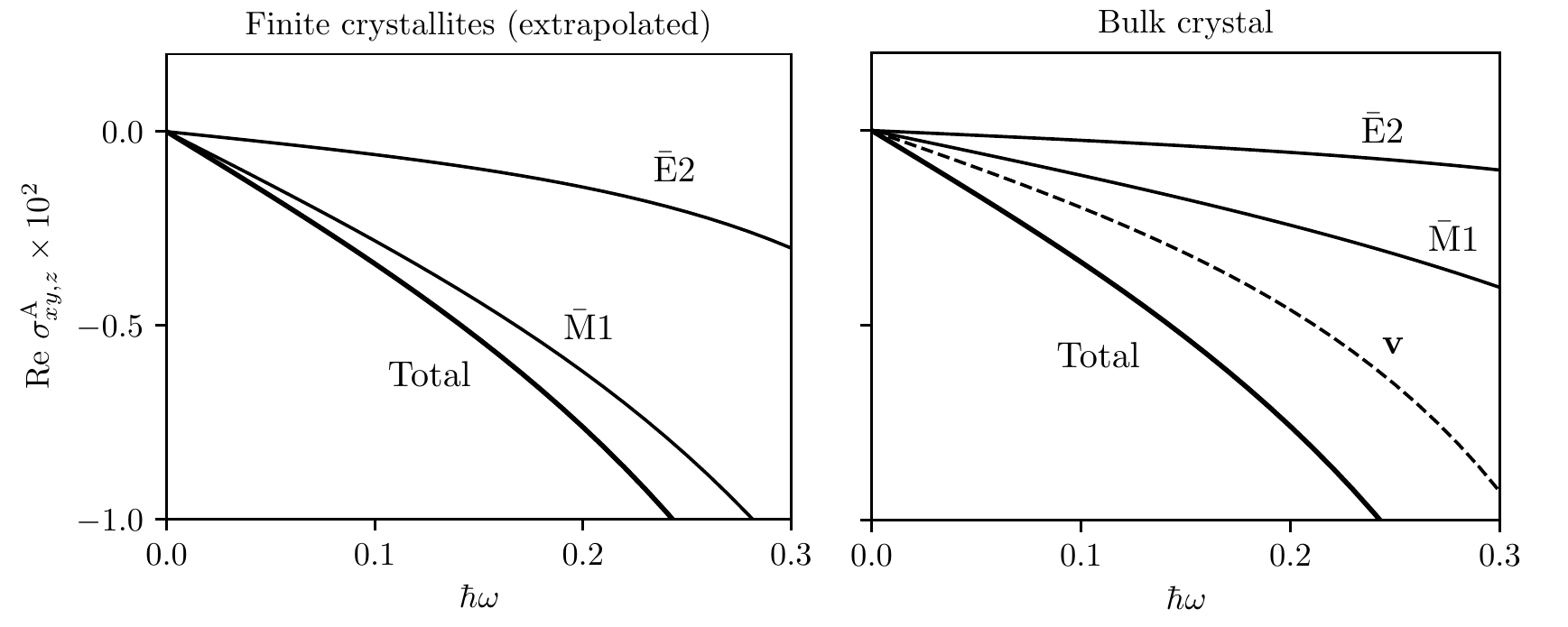}
	\caption{Decomposition of the $\sigma_{xy,z}^{\mathrm{A}}$ curves in
		\fref{fig2} into three types of origin-independent contributions:
		intrinsic magnetic dipole ($\MDbar$), intrisic electric quadrupole
		($\EQbar$), and band dispersive ($\v$). The latter is only present
		in the bulk calculation on the right, and it must be included to
		obtain the same total result as in the extrapolated crystallite
		calculations on the left.}
	\figl{fig3}
\end{figure*}

\section{Summary and discussion}
\secl{summary}

In summary, we have developed a band-theoretical description of
optical spatial dispersion in insulating and conducting crystals.  The
novelty with respect to previous formulations resides in the fact that
the current induced by the optical field is given in a
physically-transparent form, as a sum of contributions that are
individually origin independent, and which remain invariant under
single-band gauge transformations of the Bloch eigenstates.  Although
we have focused on the optical conductivity $\sigma_{ab,c}(\w)$ at
first order in $\q$, higher-order responses can in principle be
treated in a similar manner.

\begin{figure}[!t]
	\centering
	\includegraphics{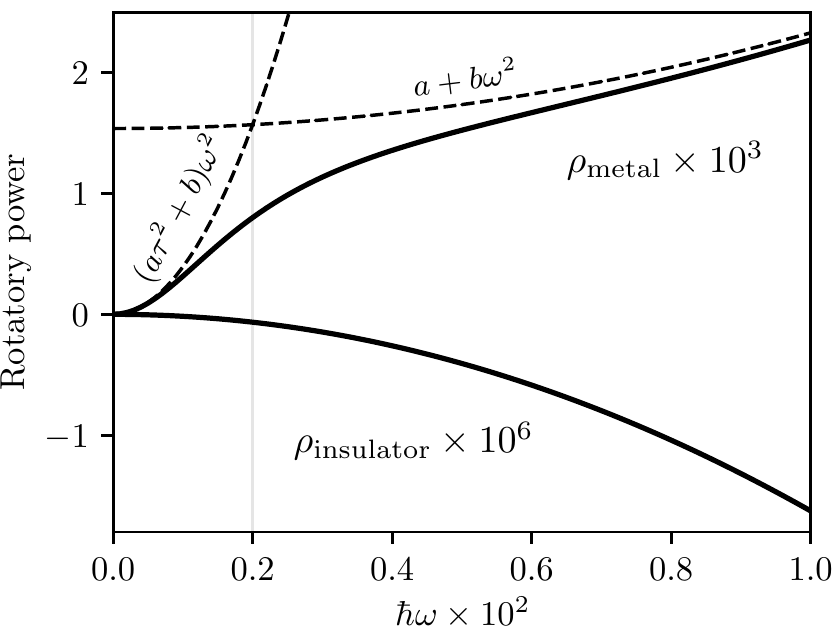}
	\caption{Low-frequency optical rotatory dispersion of the model of
		\eq{TB-model} and \fref{fig1} in an insulating state
		($\varepsilon_\text{F}=0.0$), and in a metallic state
		($\varepsilon_\text{F}=1.0$). Numerical results based on
		Eqs.~\eqref{eq:kubo-A} and \eqref{eq:rho}
		are shown as solid lines. The dashed lines are guides to the eye,
		indicating the distinct parabolic behaviors in the metallic state
		for $\w\tau\ll 1$ and for $\w\tau\gg 1$, as per
		\eq{ord-metals}, and the faint vertical line indicates the
		crossover frequency $\omega=1/\tau$. The rotatory power has units
		of radians per length.}
	\figl{fig4}
\end{figure}

For a crystal composed of nonoverlapping units, our formula for
$\sigma_{ab,c}(\w)$ reduces to the standard multipole-theory
expression for molecules, but with the transition moments $(d,m,q)$ of
\eq{E1-M1-E2} replaced by their intrinsic (origin-independent)
counterparts $(\bar d,\bar m,\bar q)$ given by
\eq{E1bar-M1bar-E2bar-mol}.  Away from the molecular limit,
$\sigma_{ab,c}(\w)$ changes in two ways. First, the intrinsic
transition moments between delocalized Bloch eigenstates are no longer
given by \eq{E1bar-M1bar-E2bar-mol}; instead, one should use either
the quantum-geometric expressions in \eq{geom}, or their
sum-over-states counterparts in \eq{sos}. The second change is that
$\sigma_{ab,c}(\w)$ acquires additional band-dispersion contributions
associated with electron transfer between crystal cells; this is in
line with the modern theories of electric polarization and orbital
magnetization in crystals~\cite{vanderbilt-book18}.

There were two key aspects to our derivation. The first was the use of
covariant Bloch-state derivatives to expand the optical conductivity
in powers of $\q$; this allowed us to eliminate spurious noncovariant
terms, and to isolate the physically relevant matrix elements
$\bar d$, $\bar m$, and $\bar q$.  The second was the choice of the
nonsingular form of the Kubo formula in \eq{Kubo-nodiv}, rather than
\eq{Kubo-VelocityGauge}, as the starting point for the expansion in
$\q$.  An order-by-order analysis of the equivalence between the Kubo
formulas~\eqref{eq:Kubo-VelocityGauge} and~\eqref{eq:Kubo-nodiv} led
us to identify a hierarchy of optical sum rules. In particular, we
found that the well-known rotatory-strength sum rule from molecular
physics remains valid for crystals, thanks to a topological argument
involving the $\kk$-space Berry curvature.

Our work opens up new prospects for realistic \textit{ab initio}
calculations of spatially-dispersive optical responses in crystals,
and of static magnetoelectric and quadrupolar responses as well.  An
implementation based on the sum-over-states formulas for
$(\bar d,\bar m,\bar q)$ in \eq{sos} has already been carried out in a
concurrent work done in coordination with the present
one~\cite{wang-prb23}, and we also envision evaluating the
$\kk$-derivative formulas in \eq{geom} using Wannier
interpolation~\cite{wang-prb06}.

In closing, we mention a possible connection with the theory of the
orbital Hall effect. It was recently
proposed~\cite{bhowal-prl21,cysne-prb22,pezo-prb22} to evaluate the
orbital Hall conductivity using an orbital-current operator whose
matrix elements in the Bloch-eigenstate basis are proportional to
$\sum_l
\big(m^{z,\text{orb}}_{ml}v^a_{ln}+v^a_{ml}m^{z,\text{orb}}_{ln}\big)$.
Here, $\m^\text{orb}_{ln}$ is the bulk generalization of \eq{M1},
given by the same expression as in \eq{M1-geom} for
$\bar\m^\text{orb}_{ln}$, but with the covariant derivatives therein
replaced by ordinary derivatives.  The two definitions are related by
\beq
\bar\m^\text{orb}_{ln}=\m^\text{orb}_{ln}-\frac{e}{4i}\w_{ln}
\left(\AA_l+\AA_n\right)\times\A_{ln}\,,
\eql{mbar-m}
\eeq
and therefore they agree for $l=n$ only.  For $l\not=n$, the two terms
on the right-hand side of \eq{mbar-m} are not separately gauge
covariant, and the lack of gauge covariance of $\m^\text{orb}_{ln}$
makes the orbital Hall conductivity gauge dependent. This suggest that
one should generally use $\bar\m^\text{orb}_{ln}$ instead of
$\m^\text{orb}_{ln}$ when evaluating the orbital Hall conductivity. In
other words, one is allowed to work with $\m^\text{orb}_{ln}$ only in
the parallel-transport gauge, where $\AA_l=\AA_n=\0$.

\section*{Acknowledgements}
 
This work was supported by the IKUR Strategy under the collaboration
agreement between the Ikerbasque Foundation and the Material Physics
Center on behalf of the Department of Education of the Basque
Government, and by Grant No.  PID2021-129035NB-I00 funded by
MCIN/AEI/10.13039/501100011033.  The authors wish to thank Cheol-Hwan
Park, Stepan Tsirkin, Xiaoming Wang, Fernando de Juan, Raffaele Resta,
and Massimiliano Stengel for valuable discussions and comments on the
manuscript.






\nolinenumbers

\end{document}